\documentclass[conference,compsoc]{IEEEtran}

\newcommand{\repoUrl}{\href{https://github.com/0xADE1A1DE/Slice-Slice-Baby}{https://github.com/0xADE1A1DE/Slice-Slice-Baby}}

\usepackage{xurl}

\newcommand{\results}[1]{%
  \ifthenelse{\equal{#1}{6700}}{1.1$\times$}{}%
  \ifthenelse{\equal{#1}{11700}}{1.5$\times$}{}%
  \ifthenelse{\equal{#1}{8700}}{8$\times$}{}%
  \ifthenelse{\equal{#1}{9850}}{9$\times$}{}%
  \ifthenelse{\equal{#1}{10900}}{10$\times$}{}%
}

\newcommand{\resultsSeconds}[1]{%
  \ifthenelse{\equal{#1}{6700}}{0.08}{}%
  \ifthenelse{\equal{#1}{11700}}{0.3}{}%
  \ifthenelse{\equal{#1}{8700}}{0.8}{}%
  \ifthenelse{\equal{#1}{9850}}{0.7}{}%
  \ifthenelse{\equal{#1}{10900}}{1.6}{}%
}

\usepackage{booktabs}
\usepackage{array}
\usepackage{caption}
\usepackage{adjustbox}
\usepackage{rotating}
\usepackage[usestackEOL]{stackengine}
\usepackage{threeparttable}
\usepackage{tabularx}
\usepackage{comment}
\usepackage{siunitx}
\usepackage{multicol}

\newcolumntype{S}{>{\raggedleft\arraybackslash}X[table-format=5.0(3)]}

\usepackage{tikz}
\usepackage{circuitikz}
\usetikzlibrary{arrows}
\usepackage{tikzsymbols}
\usepackage{graphicx}

\newlength{\ygrid}
\newlength{\xgrid}

\usepackage[numbers,sort]{natbib}

\usepackage{amsmath}
\usepackage{listings}
\usepackage{multirow}
\usepackage[normalem]{ulem}
\usepackage{algorithm}
\usepackage[noend]{algpseudocode}
\usepackage{hyperref}
\usepackage[capitalise,nameinlink,noabbrev]{cleveref}

\usepackage{amssymb}
\usepackage{pifont}
\usepackage{enumitem}

\usepackage{nasm_lang}
\usepackage{nasm_style}

\usepackage[caption=false,font=footnotesize,labelfont=sf,textfont=sf]{subfig}

\usepackage{fixltx2e}
\usepackage{dblfloatfix}

\newcommand{\cmark}{\ding{51}}
\newcommand{\xmark}{\ding{55}}

\hypersetup{
    colorlinks,
    linkcolor={red!50!black},
    citecolor={darkgreen!70!black},
urlcolor={blue!80!black}}

\definecolor{darkgreen}{rgb}{0,0.65,0}

\newcommand{\parhead}[1]{\vspace{2pt plus 1pt minus 1pt}\par\noindent\textbf{#1}\hspace{.5em plus .25em minus .25em}}

\newcommand{\attack}[1]{#1\xspace}
\newcommand{\pp}{\attack{Prime+\allowbreak Probe}}

\begin{document}
\bstctlcite{IEEEexample:BSTcontrol}
\pagestyle{plain}
\pagenumbering{gobble}

\title{\Large \textbf Slice+Slice Baby: Generating Last-Level Cache Eviction Sets in the Blink of an Eye}

\author{
    \IEEEauthorblockN{
    Bradley Morgan\IEEEauthorrefmark{1}\IEEEauthorrefmark{2},
    Gal Horowitz\IEEEauthorrefmark{3},
    Sioli O’Connell\IEEEauthorrefmark{1},
    Stephan van Schaik\IEEEauthorrefmark{4},
    Chitchanok Chuengsatiansup\IEEEauthorrefmark{5} \\
    Daniel Genkin\IEEEauthorrefmark{6},
    Olaf Maennel\IEEEauthorrefmark{1},
    Paul Montague\IEEEauthorrefmark{2},
    Eyal Ronen\IEEEauthorrefmark{3} and
    Yuval Yarom\IEEEauthorrefmark{7}
    }
\IEEEauthorblockA{\IEEEauthorrefmark{1}The University of Adelaide \IEEEauthorrefmark{2}Defence Science and Technology Group}
\IEEEauthorblockA{\IEEEauthorrefmark{3}Tel-Aviv University \IEEEauthorrefmark{4}University of Michigan \IEEEauthorrefmark{5}The University of Klagenfurt}
\IEEEauthorblockA{\IEEEauthorrefmark{6}Georgia Tech \IEEEauthorrefmark{7}Ruhr University Bochum}
}

\maketitle

\begin{abstract}
An essential step for mounting cache attacks is finding eviction sets, collections of memory locations that contend on cache space.
On Intel processors, one of the main challenges for identifying contending addresses is the sliced cache design, where the processor hashes the physical address to determine where in the cache a memory location is stored.
While past works have demonstrated that the hash function can be reversed, they also showed that it depends on physical address bits that the adversary does not know.

In this work, we make three main contributions to the art of finding eviction sets.
We first exploit microarchitectural races to compare memory access times and identify the cache slice to which an address maps.
We then use the known hash function to both reduce the error rate in our slice identification method and to reduce the work by extrapolating slice mappings to untested memory addresses.
Finally, we show how to propagate information on eviction sets across different page offsets for the hitherto unexplored case of non-linear hash functions.

Our contributions allow for entire LLC eviction set generation in \resultsSeconds{9850} seconds on the Intel i7-9850H and \resultsSeconds{10900} seconds on the i9-10900K, both using non-linear functions.
This represents a significant improvement compared to state-of-the-art techniques taking \results{9850} and \results{10900} longer, respectively.
\end{abstract}
\section{Introduction}\label{sec:introduction}
In the two decades since their introduction~\cite{tsunooCryptanalysisBlockCiphers2002, tsunooCryptanalysisImplementedComputers2003, osvikCacheAttacksCountermeasures2005, percivalCacheMissingFun2005}, cache attacks have been recognised as a significant security threat.
Multiple attacks have been published, leaking secret or sensitive information from a wide range of applications, including 
cryptography~\cite{yaromFLUSHRELOADHigh2014, grussFlushFlushFast2016, liuLastLevelCacheSideChannel2015, horowitzSpecoScopeCacheProbing2024, yanAttackDirectoriesNot2019},
user interface~\cite{294651, 272258, grussFlushFlushFast2016},
and others~\cite{294601, orenSpySandboxPractical2015, 244042, 277232}.
Moreover, in recent years, cache attack techniques are being used as stepping stones for more advanced attacks~\cite{kocherSpectreAttacksExploiting2019, lippMeltdownReadingKernel2018, 10.1007/978-3-319-40667-1_15}.

A fundamental prerequisite for many contention-based cache attacks is to find an \emph{eviction set}, a collection of memory addresses that contend on cache space with a victim memory address.
In a typical \pp attack~\cite{osvikCacheAttacksCountermeasures2005, liuLastLevelCacheSideChannel2015}, the eviction set serves two purposes.
First, accessing the eviction set creates contention, forcing eviction of the victim memory address.
Second, by measuring the access time to memory addresses in the eviction set, the attacker can determine whether they are cached and from that infer whether the victim has accessed the victim memory address.

The main challenge for constructing eviction sets is \emph{addressing uncertainty}~\cite{WuXW12}, which hides cache addressing information from the attacker.
Standard threat models involve attackers executing unprivileged user-level code under virtual memory addressing.
Conversely, caches typically use the physical memory address for determining the cache location in which memory is stored.
The virtual-to-physical address mapping hides the physical address from the attacker, hampering direct determination of cache location from virtual addresses.

Early works overcome addressing uncertainty by either targeting the L1 cache~\cite{osvikCacheAttacksCountermeasures2005, percivalCacheMissingFun2005}, where the address bits that determine the cache location are preserved by the virtual to physical address translation or by using huge pages to ensure that more address bits are preserved~\cite{irazoquiSharedCacheAttack2015}.
However, the sliced design of the Intel last-level cache (LLC) exacerbates addressing uncertainty and precludes such techniques from working for this cache on more recent processors.
The Intel sliced cache is partitioned into multiple slices, each serving a different part of the physical address space.
To determine the slice that serves a memory address, the processor computes an undisclosed hash function based on the physical address bits.
Past efforts for reverse engineering the slice function have demonstrated that it uses all of the address bits~\cite{hundPracticalTimingSide2013, yaromMappingIntelLastLevel2015, mccalpinMappingAddressesL32021, mauriceReverseEngineeringIntel2015, irazoquiSystematicReverseEngineering2015}.
Consequently, without knowing the full physical address, the adversary cannot determine the slice to which a memory location maps.

\citet{liuLastLevelCacheSideChannel2015} propose to exploit cache misses due to contention in order to construct eviction sets.
This approach has been improved through a series of algorithmic~\cite{vilaTheoryPracticeFinding2019, kessousPrune+PlumTreeFindingEviction2024, songDynamicallyFindingMinimal2019, xueCTPPFastStealth2023} and technical~\cite{zhaoLastLevelCacheSideChannel2024, purnalPrimeScopeOvercoming2021, ThomaG22} improvements.
Due to their importance for cache attacks, the difficulty of constructing eviction sets is considered a measure for the security of defences against side-channel attacks~\cite{WernerUG0GM19, Qureshi18, GenkinKLTUY23}.
In response, some designs that are capable of finding eviction sets in the presence of side-channel countermeasures have been proposed~\cite{PurnalBPV23, PurnalGGV21, katzmanGatesTimeImproving2023}.

Most proposed approaches for eviction-set creation exploit the observation that when the number of slices is a power of two, eviction sets at different page offsets are equivalent~\cite{liuLastLevelCacheSideChannel2015, yaromMappingIntelLastLevel2015}.
Thus, on such machines, the attacker only needs to find eviction sets for a single page offset and can directly propagate that information to all other page offsets in the page.
This approach does not work for machines where the number of slices is not a power of two and the slice function is not linear~\cite{yaromMappingIntelLastLevel2015}.
Nonetheless, \citet{yaromMappingIntelLastLevel2015} demonstrate that for a six-core machine, it may be possible to propagate information from some page offsets to the rest of the page.

\subsection{Our Contribution}\label{subsec:our-contribution}
In this work, we present further improvements to the art of eviction set construction.
For that we build on two main advancements.
First, we show how to use timing variations to determine the slice a memory address maps to.
This information allows us to further partition the initial candidate set memory, achieving additional speedup.
Secondly, we show how to use properties of the known hash function to carry over information across page offsets, allowing us to recover eviction sets for the whole address space while only probing a fraction of the cache lines.
We now describe our algorithm in further detail.

As a first step for our algorithm, we develop a technique to identify the cache slice that a memory location maps to by exploiting differences in latency that each slice exhibits due to the processor's physical layout.
The main challenge is that due to system load and changes in processor frequency, the latency of slice accesses vary and overlap with one another, making it difficult to narrow down the slice that a memory location maps to.
To overcome this challenge, we build on microarchitectural weird machines~\cite{evtyushkinComputingTimeMicroarchitectural2021, katzmanGatesTimeImproving2023, kaplanOptimizationAmplificationCache2023, horowitzSpecoScopeCacheProbing2024, WangPWB24}, constructing a gadget which uses a memory access race condition to determine that with the lower access time.
We refer to this as the comparator gate, and observe that because system load impacts memory accesses similarly, comparing these helps lessen the effects of noise.
Using our comparator gate, we determine the slice that a memory location maps to with an average accuracy of 97\% from any processor core.

Then, we propose an approach to improve the speed of our algorithm by extrapolating slice mappings from a few recovered memory addresses to other addresses in the same page.
Specifically, we exploit the observation that the processor's hash function generates only a few possible mappings from virtual page offsets to slices.
Consequently, by recovering a few locations in a virtual page, we can determine its overall slice mapping and propagate this to the rest of the locations in the page, greatly speeding up an eviction set construction.
Moreover, this technique also allows for error detection, because in the case of error, there is a high probability that the recovered slice numbers do not match any of the possible patterns.

To construct eviction sets, we generate a set of candidate addresses, all of which reside in the same page offset.
We partition this set, first based on the L2 cache set~\cite{zhaoLastLevelCacheSideChannel2024} and secondly based on the LLC slice as determined by our comparator gate and slice mapping propagation method.
We then use a conflict-based eviction-set construction algorithm on each partition separately.

As a final contribution, we show how to exploit the knowledge about the hash function used for mapping memory into slices to propagate eviction set information on non-linear slice functions.
Specifically, we demonstrate that by constructing eviction sets in 4\,KB page offsets, we can identify working eviction sets in other offsets, requiring only 15--22\% of the total eviction sets to be built conventionally, depending on the function.
This achieves a total execution time improvement of \results{8700} to \results{10900} over prior work.
For machines with linear slice functions, our algorithm is \results{6700} to \results{11700} faster than state-of-the-art techniques.

In summary, the contributions of this work are:
\begin{itemize}[nosep,left=0pt]
\item We demonstrate a new weird gate construction that can identify the LLC slice that a memory location maps to~(\cref{sec:transient-address-access-timing}).
\item We show how to use the known hash function to detect and correct errors in slice mapping as well as to propagate slice information across pages~(\cref{sec:slice-decision-tree}).
\item We demonstrate how to propagate eviction-set information across page offsets when the processor uses a non-linear hash function~(\cref{sec:slice-aware-eviction-set-creation}).
\end{itemize}

\smallskip
Following the practices of open science, we make our source code and research artifacts publicly available at \repoUrl.

\section{Background and Related Work}\label{sec:background}
In this section, we introduce the necessary background on Intel caches, their LLC slice function and past recovery efforts, eviction sets and microarchitectural weird gates.

\subsection{Intel Cache Hierarchy}\label{subsec:intel-cache-heirarchy}
Processor caches store sections of main memory to reduce access time.
A cache miss occurs when the requested data is absent in the cache, whereas a hit occurs when the data is found.
Modern caches are typically $n$-way set-associative, split into $m$-sets with specific indexing from memory address bits.
Each Intel CPU core has its own L1 data and instruction caches, and also an L2 cache.
The last-level cache (LLC) is the largest, shared across cores and located in the uncore section of the processor die~\cite{Intel64IA32}, containing all non-CPU core logic.
Accessing memory loads a 64-byte chunk into the L1 cache as a cache line, with evictions occurring in accordance with replacement policies from L1 to L2 and beyond.
Older Intel CPUs feature an \emph{inclusive} LLC, holding copies of L1 and L2 data\@.
Modern processors use \emph{non-inclusive} caches~\cite{intel12thGenerationIntel}, where data held in the private caches may or may not appear in the LLC\@.
Here, a coherency directory (CD) or snoop filter (SF) is used to maintain cache coherency without storing duplicate data from lower-level caches~\cite{Intel64IA32, intelIntelXeonProcessor}.
These mechanisms track the state and location of cached data across cores and have a similar set-associative structure as the caches~\cite{yanAttackDirectoriesNot2019, zhaoLastLevelCacheSideChannel2024}.
For our work, we focus on several Intel processors with inclusive and non-inclusive LLCs.
\cref{tab:cache-parameters} shows the cache parameters for such processors.

\begin{table}[t]
\caption{Intel cache parameters with $m$-sets and $n$-ways.}
\begin{adjustbox}{max width=\linewidth}
\begin{tabular}{lrrrrrrrr}
\toprule
\textbf{}                                                                       & \multicolumn{2}{c}{\textbf{L1 Data}} & \multicolumn{2}{c}{\textbf{L1 Instr.}} & \multicolumn{2}{c}{\textbf{L2}} & \multicolumn{2}{c}{\textbf{LLC / Slice}} \\ \cmidrule(l){2-3} \cmidrule(l){4-5} \cmidrule(l){6-7}\cmidrule(l){8-9}
\textbf{Processor}                                                              & \multicolumn{1}{c}{$m$} & \multicolumn{1}{c}{$n$} & \multicolumn{1}{c}{$m$} & $n$ & \multicolumn{1}{c}{$m$} & \multicolumn{1}{c}{$n$} & \multicolumn{1}{c}{$m$} & $n$      \\
\midrule
\begin{tabular}[c]{@{}l@{}}i7-6700K, i7-8700\\ i7-9850H, i9-10900K\end{tabular} & 64                & 8                & 64                 & 8                 & 1024           & 4              & 1024                & 16                 \\
i7-11700K                                                                       & 64                & 12               & 64                 & 8                 & 1024           & 8              & 1024                & 16                 \\
\begin{tabular}[c]{@{}l@{}}i7-12900KF (P-Core)\\i7-13700H (P-Core)\end{tabular} & 64                & 12               & 64                 & 8                 & 2048           & 10             & 4096                & 12                 \\
i9-14900K (P-Core)                                                              & 64                & 12               & 64                 & 8                 & 2048           & 16             & 4096                & 12                 \\
\bottomrule
\end{tabular}
\end{adjustbox}
\label{tab:cache-parameters}
\end{table}

\subsection{Sliced Caches}\label{subsec:sliced-caches}
\begin{figure}[!t]
\footnotesize
\begin{center}
\begin{adjustbox}{height=4.2cm}
\begin{circuitikz}
    \tikzstyle{every node}=[font=\Large]
    \draw [ fill={rgb,255:red,255; green,239; blue,202} ] (11.25,14.00) rectangle  node {\large LLC Slice 0} (13.75,13.00);
    \draw [ fill={rgb,255:red,255; green,239; blue,202} ] (14.5,14.00) rectangle  node {\large LLC Slice 1} (17,13.00);
    \draw [ fill={rgb,255:red,255; green,239; blue,202} ] (11.25,12.25) rectangle  node {\large LLC Slice 3} (13.75,11.25);
    \draw [ fill={rgb,255:red,255; green,239; blue,202} ] (14.5,12.25) rectangle  node {\large LLC Slice 2} (17,11.25);
    \draw [ fill={rgb,255:red,218; green,255; blue,208} , rounded corners = 16.2] (11.25,11) rectangle  node {\large Core 3} (13.75,9.75);
    \draw [ fill={rgb,255:red,218; green,255; blue,208} , rounded corners = 16.2] (14.5,11) rectangle  node {\large Core 2} (17,9.75);
    \draw [ fill={rgb,255:red,218; green,255; blue,208} , rounded corners = 16.2] (11.25,15.5) rectangle  node {\large Core 0} (13.75,14.25);
    \draw [ fill={rgb,255:red,218; green,255; blue,208} , rounded corners = 16.2] (14.5,15.5) rectangle  node {\large Core 1} (17,14.25);
    \draw [line width=2pt, ->, >=Stealth] (10.25,12.5) -- (10.25,12.75);
    \draw [line width=2pt, ->, >=Stealth] (18,12.75) -- (18,12.5);
    \draw [line width=2pt, ->, >=Stealth] (14,14.25) -- (14.25,14.25);
    \draw [line width=2pt, ->, >=Stealth] (14.25,11) -- (14,11);
    \draw [ line width=1.6pt , rounded corners = 26.4] (10.25,14.25) rectangle  node {\large Uncore Ring Interconnect} (18,11);

    \draw [line width=2pt, ->, >=Stealth] (10.50,12.75) -- (10.50,12.5);
    \draw [line width=2pt, ->, >=Stealth] (17.75,12.5) -- (17.75,12.75);
    \draw [line width=2pt, ->, >=Stealth] (14.25,14) -- (14,14);
    \draw [line width=2pt, ->, >=Stealth] (14,11.25) -- (14.25,11.25);
    \draw [ line width=1.6pt , rounded corners = 20.0] (10.50,14.00) rectangle (17.75,11.25);

    \draw [dash pattern=on 2pt off 2pt, color=gray, line width=0.5mm, opacity=0.6] (12.5,14.00) -- (12.5,13.00);
    \draw [dash pattern=on 2pt off 2pt, color=gray, line width=0.5mm, opacity=0.6] (15.75,14.00) -- (15.75,13.00);
    \draw [dash pattern=on 2pt off 2pt, color=gray, line width=0.5mm, opacity=0.6] (12.5,12.25) -- (12.5,11.25);
    \draw [dash pattern=on 2pt off 2pt, color=gray, line width=0.5mm, opacity=0.6] (15.75,12.25) -- (15.75,11.25);

    \draw[overlay] node at (12.5, 13.5) {\large LLC Slice 0};
    \draw[overlay] node at (15.75, 13.5) {\large LLC Slice 1};
    \draw[overlay] node at (15.75, 11.75) {\large LLC Slice 2};
    \draw[overlay] node at (12.5, 11.75) {\large LLC Slice 3};
\end{circuitikz}
\end{adjustbox}
\end{center}
\caption{Logical structure of the ring interconnect on Intel Core CPUs. It allows for the bi-directional data transfer~\cite{paccagnellaLordRingSide, 10.1007/978-3-030-80825-9_14} between cores, slices/sub-slices and other structures.}
\label{fig:core-uncore}
\end{figure}
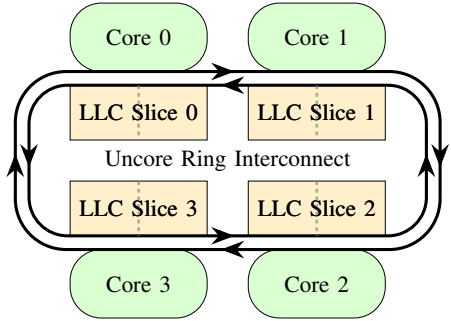

Intel's design of the LLC splits it into several distinct \emph{slices}, separately connected to the processor's interconnect in the uncore region of the processor~\cite{intel6thGenerationIntel2016, intelIntelXeonProcessor, 10.1007/978-3-030-80825-9_14, paccagnellaLordRingSide}.
Memory maps to certain slices based on the physical address according to a proprietary hash function undisclosed by Intel.
Based on prior reverse engineering efforts~\cite{hundPracticalTimingSide2013, irazoquiSystematicReverseEngineering2015, mauriceReverseEngineeringIntel2015, inciSeriouslyGetMy2015, yaromMappingIntelLastLevel2015, mccalpinMappingAddressesL32021}, each of the slices are attached to hardware structures known as a C-Box or CBo, one per CPU core~\cite{intel6thGenerationIntel2016}.
From experimentation with eviction sets~\cite{vilaTheoryPracticeFinding2019} on Intel processors with inclusive LLCs, the set mappings of the eviction addresses imply the existence of double the number of slices than reported by the processor.\footnote{As reported by reading Model Specific Register \texttt{0x396}.}
This can be characterised as each slice being split into two sub-slices per CBo, as the dashed lines in \cref{fig:core-uncore} illustrate.

\parhead{Linear Functions.}
When the number of LLC slices is a power of two, the processor calculates the slice index bits using a straightforward linear XOR operation on the physical address bits.
This technique employs a set of processor model-specific permutation masks to select appropriate address bits and determine the slice index~\cite{mauriceReverseEngineeringIntel2015, mccalpinMappingAddressesL32021}.

\parhead{Non-Linear Functions.}
When the number of LLC slices is not a power of two, the employed function is not linear.
Instead, it consists of two phases, a linear XOR permutation selection similar to the linear functions, piped into a secondary stage which outputs slice values in the appropriate range.
\citet{yaromMappingIntelLastLevel2015} describe this secondary step, which can either be represented as a logic circuit or a lookup of mappings for the slice values termed a \emph{base sequence} by \citet{mccalpinMappingAddressesL32021}.
The base sequence is a sequence of slice values which are indexed by the output of the linear XOR operation.
In this work we treat the secondary function as such a sequence.

\subsection{LLC Contention-Based Cache Attacks}\label{subsec:prime+probe-style-cache-attacks}
The last-level cache (LLC) has become a compelling target for contention-based attacks as it is shared across all CPU cores.
It permits information leakage between cores of the same physical processor, with several prior attacks exploring this possibility~\cite{liuLastLevelCacheSideChannel2015, younisNewPrimeProbe2015, disselkoenPrimeAbortTimerFree2017, yanAttackDirectoriesNot2019, purnalPrimeScopeOvercoming2021, zhaoLastLevelCacheSideChannel2024}.
In these cross-core attacks, an attacker first primes the shared cache with their own memory using \emph{eviction sets} to move the victim's data into RAM\@.
The attacker then waits for the victim to run and displace the eviction set memory.
Through, e.g.\ observation of the access times to their own data, an attacker can infer the victim's memory accesses, allowing for the recovery of sensitive information such as cryptographic secret keys~\cite{10.1007/978-3-319-93387-0_5, liuLastLevelCacheSideChannel2015, younisNewPrimeProbe2015, irazoquiSharedCacheAttack2015} and user interface interactions~\cite{294651, 272258}.

\subsubsection{Attack Workflow}\label{subsubsec:eviction-sets}
The main goal of the attacker is to leak and recover secret data from other CPU cores.
To do this, the attacker needs to build many minimal eviction sets~\cite{vilaTheoryPracticeFinding2019, songDynamicallyFindingMinimal2019}, i.e.\ the smallest set of \emph{congruent} addresses which evict a target address into the next cache level or to RAM\@.
Two addresses are described as \emph{congruent} when they both map to the same cache set as each other.
Given an $n$-way set associative cache, a minimal eviction set contains exactly $n$ addresses.
After this, the attacker needs to narrow down which eviction sets correspond with the victim's memory to observe their behaviour~\cite{10.1007/978-3-642-10366-7_39, grussCacheTemplateAttacks2015, liuLastLevelCacheSideChannel2015, zhaoLastLevelCacheSideChannel2024}.

For an inclusive LLC, the attacker can evict the victim's memory out of the LLC (and hence the private caches) to observe memory access patterns~\cite{vilaTheoryPracticeFinding2019, liuLastLevelCacheSideChannel2015}.
However, for non-inclusive LLCs, the coherency directory (CD) or snoop filter (SF), as determined by the processor family or configuration settings~\cite{Intel64IA32, intelIntelXeonProcessor, vimalmReTurningCache2022}, provides an accessible structure to cause private cache evictions~\cite{yanAttackDirectoriesNot2019, zhaoLastLevelCacheSideChannel2024}.
These hardware structures are a finite resource tracking the locations of cached data in the LLC\@.
Thus, a congruent set of addresses which map to the same CD or SF set can likewise cause evictions into RAM\@.

\subsubsection{Finding Eviction Sets}\label{subsubsec:finding-eviction-sets}
One of the main challenges in cache attacks is finding minimal eviction sets in a virtual memory environment.
With virtually addressed 4\,KB memory pages, the attacker does not know the higher order bits of the physical address which decide the cache set index for the L2 and LLC\@.
Moreover, these bits are also required for the computation of the slice index with the hash function.
Therefore, the attacker cannot directly create eviction sets for these caches.
To find a minimal eviction set, the attacker instead needs to find an eviction set by  testing for contention.
The attacker initialises a large candidate address pool which evicts the target address from the desired cache and then reduces this pool into a minimal eviction set with a chosen pruning algorithm~\cite{vilaTheoryPracticeFinding2019, purnalPrimeScopeOvercoming2021, songDynamicallyFindingMinimal2019, zhaoLastLevelCacheSideChannel2024, kessousPrune+PlumTreeFindingEviction2024}.
These algorithms require test functions to determine whether a candidate set can actually evict a target address, with differing techniques for inclusive caches~\cite{liuLastLevelCacheSideChannel2015} and non-inclusive caches~\cite{yanAttackDirectoriesNot2019}.

\parhead{Group Testing.}
\citet{vilaTheoryPracticeFinding2019} describe the group testing method for finding minimal eviction sets for the LLC\@.
Here, the attacker splits the candidate set into $n+1$ groups (with~$n$ being the associativity of the cache) and tests the first~$n$ groups together for eviction of the target address.
If the first~$n$ groups successfully evict the address, then the last group can be pruned away, with the process repeated on now the smaller candidate set.
If the target address is not evicted, then the attacker splits away a different group and tries again.

\parhead{Prime+Scope.}
The Prime+Scope eviction set construction method for the LLC~\cite{purnalPrimeScopeOvercoming2021} works by first accessing the target address, then repeatedly guessing and accessing further addresses which may be congruent.
After each guess, the attacker checks if the target address had a fast access or not.
If it was fast, then the guessed address is appended to the eviction set.
This entire process repeats until a minimal eviction set of size $n$ is built.

\parhead{Binary Search.}
In binary search pruning~\cite{zhaoLastLevelCacheSideChannel2024}, the attacker incrementally expands a growing window of addresses to test, checking them in sequence.
The window is expanded until it contains enough congruent addresses to evict the target.
Once an eviction is confirmed, the last address in the window is moved to the front, and the attacker continues searching.
When the first $n$ addresses are confirmed to be congruent, a minimal eviction set has been identified.

\parhead{Prune+PlumTree.}
The Prune+PlumTree algorithm aims to find many independent eviction sets simultaneously from an initial candidate set~\cite{kessousPrune+PlumTreeFindingEviction2024}.
An initial pruning step identifies cached memory by accessing the entire candidate set, discarding memory which experiences a cache miss.
In the second stage, called PlumTree, the pruned candidates are recursively organised into minimal eviction sets by imposing a tree structure on the candidates.
This partitioning allows the algorithm to identify possible minimal eviction sets at the leaf nodes, discarding those at leaves which cannot evict other memory.

\subsubsection{Optimisation Techniques}
More advanced techniques can improve the efficiency of eviction set generation, one such example being conflict set reduction~\cite{liuLastLevelCacheSideChannel2015, vilaTheoryPracticeFinding2019}.
This conflict set is the union of all minimal eviction sets for the LLC, and is then split into individual eviction sets.
It acts to filter the initially large candidate set by reducing the number of addresses to test, speeding up the overall process.

Another improvement is the concept of L2 candidate set filtering~\cite{zhaoLastLevelCacheSideChannel2024}.
This technique works by reducing the number of addresses to test by filtering out those which do not correspond to the same L2 set as the target address using an L2 eviction set.
In the same work, the authors also propose using a parallel access method to reduce the time taken to test the eviction sets, reducing false positives when testing for eviction.

Previous works~\cite{liuLastLevelCacheSideChannel2015, yaromMappingIntelLastLevel2015} exploit a characteristic of linear slice functions to \emph{propagate} eviction sets from one offset to all offsets within a page.
On machines with this type of hash function, adjusting the offset bits of each address in a single eviction set consistently maps them to the same LLC slice and set across all page offsets.

\subsection{Weird Gates}\label{subsec:speculative-execution-and-weird-gates}
To reduce stalls and delays, modern processors may execute instructions out of order.
For that, the processor tracks the data dependencies of instructions and aims to execute instructions as soon as their dependencies are satisfied, even if preceding older instructions have not completed execution.
To increase the benefit of out-of-order execution, processors try to predict the outcome of control-flow instructions, such as branches, and execute instructions along the predicted path even before the processor determines the outcome of the branch.
In the case of correct prediction, this allows the processor to progress execution beyond the branch.
Conversely, if the process eventually determines that the prediction was incorrect, the processor squashes the mispredicted execution path and resumes execution from the correct branch outcome.

Squashing speculative execution does not undo the effects that the squashed instructions have on the microarchitectural state of the processor.
Weird gates~\cite{kaplanOptimizationAmplificationCache2023, katzmanGatesTimeImproving2023} exploit this observation to compute various functions on microarchitectural state.
Specifically, a typical weird gate consists of two or more \emph{instruction chains}~\cite{horowitzSpecoScopeCacheProbing2024}, possibly non-consecutive sequences of instructions, such that each instruction in the chain has a data dependency on earlier instructions in the chain.
The last instruction in a \emph{control chain} is a mispredicted control-flow instruction, whose outcome depends on executing all of the instructions in the chain.
Examples of such mispredicted control-flow instruction include conditional branches~\cite{katzmanGatesTimeImproving2023} and return instructions~\cite{kaplanOptimizationAmplificationCache2023}.

Another type of chain is a \emph{signal chain}.
Such a chain contains one or more ``signal'' instructions along the mispredicted path, whose execution leaves observable traces in the microarchitectural state of the processors.
A common example of such signal instructions is memory accesses that, when executed, result in bringing the accessed memory to the cache.

This design of weird gates creates an observable race between the signal and the control chains.
Specifically, when the execution reaches the end of the signal chain, the processor determines that a misprediction has occurred and terminates the speculative execution of the signal chain.
If the speculative execution reaches a signal instruction before it is squashed, the instruction will leave its trace in the microarchitecture.
Conversely, if the speculative execution is squashed before executing the signal instruction, the instruction will not be executed, leaving no trace.

\parhead{Weird NOT Gate.}
As a concrete example, we now describe a weird NOT gate~\cite{kaplanOptimizationAmplificationCache2023, katzmanGatesTimeImproving2023}.
The gate is implemented as a function that takes two memory addresses, \texttt{input} and \texttt{output}.
We assume that on invocation the \texttt{output} address is not cached, but make no assumption on the caching state of the \texttt{input} address.
The aim of the gate is that after its execution, \texttt{output} will be in the cache if and only if originally \texttt{input} was not in the cache.

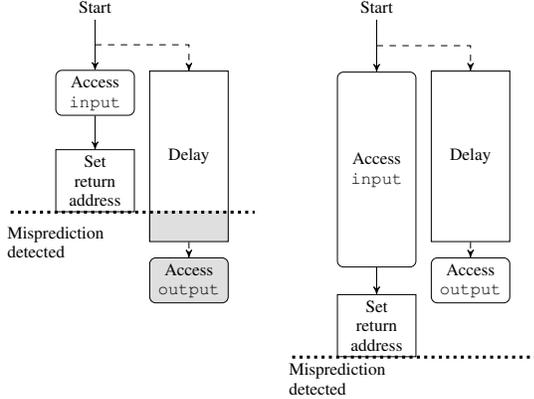
\begin{figure}[!t]
\footnotesize
\setlength{\ygrid}{2cm}
\setlength{\xgrid}{2.2cm}
\begin{center}
\adjustbox{width=0.85\linewidth}{
\begin{tikzpicture}[-stealth', scale=0.7]
  \tikzstyle{textbox} = [rectangle, minimum height=.5cm,text width=0.5\xgrid, minimum width=0.5\xgrid, text centered, draw=black, rounded corners =.1cm]
  \tikzstyle{chain} = [textbox, rounded corners=0cm]
  \tikzstyle{nospec} = [fill=lightgray!50!white]
  \node(FENCE){Start};
  \node[textbox] at (0,-\ygrid) (IN){Access \texttt{input}};
  \node[textbox, chain, anchor=south] at (0,-2.4\ygrid) (RETADR) {\footnotesize Set return address};
  \node[minimum height=0.25\ygrid, minimum width=0.6\xgrid, nospec, anchor=south] at (\xgrid, -2.75\ygrid) {};
  \draw[-, dotted, line width=.5mm] (-.9\xgrid, -2.4\ygrid) -- (1.7\xgrid, -2.4\ygrid);
  \node[anchor=north west, text width=1\xgrid] at (-\xgrid, -2.5\ygrid) {\footnotesize Misprediction\\detected};
  \draw (FENCE) --coordinate[midway](S1P) (IN);
  \draw (IN)-- (RETADR);
  \node[chain, minimum height=1.4\ygrid, anchor=south] at (1\xgrid,-2.75\ygrid) (STR1) {Delay};
  \draw[dashed] (S1P) -|  (STR1);
  \node[textbox, nospec] at (1\xgrid,-3.2\ygrid) (SIG1) {Access \texttt{output}};
  \draw[dashed] (STR1) -- (SIG1);

  \begin{scope}[xshift=3\xgrid]
    \node(FENCE){Start};
    \node[textbox, minimum height=1.6\ygrid] at (0,-1.9\ygrid) (IN){Access \texttt{input}};
    \node[textbox, chain, anchor=south] at (0,-4.1\ygrid) (RETADR) {\footnotesize Set return address};
    \draw[-, dotted, line width=.5mm] (-.9\xgrid, -4.1\ygrid) -- (1.7\xgrid, -4.1\ygrid);
    \node[anchor=north west, text width=1\xgrid] at (-\xgrid, -4.1\ygrid) {\footnotesize Misprediction\\detected};
    \draw (FENCE) --coordinate[midway](S1P) (IN);
    \draw (IN)-- (RETADR);
    \node[chain, minimum height=1.4\ygrid, anchor=south] at (1\xgrid,-2.75\ygrid) (STR1) {Delay};
    \draw[dashed] (S1P) -|  (STR1);
    \node[textbox] at (1\xgrid,-3.2\ygrid) (SIG1) {Access \texttt{output}};
    \draw[dashed] (STR1) -- (SIG1);

  \end{scope}
\end{tikzpicture}
}
\caption{Operation of the  weird NOT gate, which inverts the cached state of \texttt{input} to \texttt{output}. (Adapted from: \citet{horowitzSpecoScopeCacheProbing2024}.)}
\label{fig:not-gate}
\end{center}
\end{figure}

\cref{fig:not-gate} illustrates the operation of the gate in the case that \texttt{input} is in the cached (left) and when it is not (right).
The signal chain consists of a delay, generated by a sequence of arithmetic instructions, followed by an access to \texttt{output}.
The control chain consists of an access to \texttt{input}, followed by updating the return address so that executing the return instruction skips the execution of the signal chain.

When \texttt{input} is in the cache (\cref{fig:not-gate} left), retrieving its value is fast and the control chain completes before the control chain accesses \texttt{output}.
Hence, in this case, \texttt{output} remains uncached.
Conversely, if \texttt{input} is not in the cache, accessing it will be longer, delaying the detection that the return is mispredicted.
This would allow the signal chain to access \texttt{output}, bringing its contents to the cache.

\section{Overview}\label{sec:overview}
The primary goal of our work is to increase the feasibility of contention-based cache attacks by speeding up LLC eviction set generation, the initial step in contention-based cross-core cache attacks.
The main challenge we overcome is the uncertainty posed by the LLC slice function.
We achieve that through our slice prediction technique in \cref{sec:transient-address-access-timing}, where we group memory by slice based on observed microarchitectural effects using a weird gate for memory access time comparison.

We further enhance our slice-mapping approach in \cref{sec:slice-decision-tree} by propagating fewer slice guesses across entire 4\,KB memory pages using knowledge of the processor's slice function, accelerating the memory classification speed while additionally reducing errors.
This allows us to infer nearby address slices without needing to measure them directly by finding the closest matching whole-page slice permutation.

Finally, we bring these efforts together to improve LLC eviction set generation with slice-aware optimisations in \cref{sec:slice-aware-eviction-set-creation}.
We combine our approach with state-of-the-art L2 candidate set filtering by \citet{zhaoLastLevelCacheSideChannel2024}, comparing to their work as well as the Prune+PlumTree algorithm~\cite{kessousPrune+PlumTreeFindingEviction2024}.

\parhead{Threat Model.}
Our work follows the standard cross-core attack threat model where the adversary can only execute \textit{unprivileged} code on the same physical processor as the victim, but not on the same physical core.
We assume a sliced last-level cache, as used in modern Intel processors, with a known slice function (having recovered it offline, a one-time reverse engineering effort).
We do not require the use of huge pages, relying solely on standard 4\,KB memory pages (i.e.\ we do not have visibility of the higher-order bits for the addresses of memory we use).
Moreover, we do not disable any of the hardware prefetchers.

\section{Determining Slice Mappings for Memory}\label{sec:transient-address-access-timing}
In this section, we demonstrate our methodology for determining LLC slice indices for memory without requiring root access.
Our eviction set generation routine leverages the ability to group memory addresses based on their slice mappings to gain several speed improvements by exploiting observable differences in access latency to the LLC\@.
We experiment with several options to predict slice mappings for memory, exploring the use of the RDTSCP instruction as well as a variant of a weird NOT gate to measure LLC access times.
However, we find that under high system noise both approaches fail to accurately predict the LLC slice of memory addresses.

We then introduce the \emph{comparator} gate, a new type of weird gate that compares access timing of two memory addresses, instead of using a fixed delay, as in prior weird gates designs.
We find that this comparison approach is less sensitive to noise than the other methods and that it consistently maintains high classification accuracy, regardless of which core we measure from.

\parhead{L2 Eviction Sets.}
Throughout this section, we require the use of L2 eviction sets to place memory addresses in the LLC to then measure access times using several techniques.
An L2 eviction set consists of a set of memory addresses that, when accessed, evict a target address from the private caches into the shared LLC, enabling our experiments.
Generating L2 eviction set only takes tens of milliseconds (depending on the processor) due to the small  size in comparison to the LLC and the lack of slice function.
We use group testing~\cite{vilaTheoryPracticeFinding2019} for L2 eviction set creation
and use small 4\,KB memory pages when creating to mimic the scenario of a real-world system.

\subsection{Predicting Slices with Access Time}\label{subsec:llc-slice-access-latency}
The sliced LLC implemented by Intel exhibits different access times depending on the slice a memory address belongs to~\cite{doweck6thGenerationIntelCore2017}.
This has been explored in previous works as a method to reverse-engineer the slice function~\cite{yaromMappingIntelLastLevel2015}, as well as to create covert communication side-channels and mount cross-core attacks~\cite{paccagnellaLordRingSide, 10.1007/978-3-030-80825-9_14}.
As visualised in \cref{fig:core-uncore}, cores with memory residing in distant slices must wait for their traffic to travel further than if the memory resides in a closer slice.

\begin{figure}[!t]
\begin{center}
\begin{adjustbox}{width=\linewidth}
\includegraphics{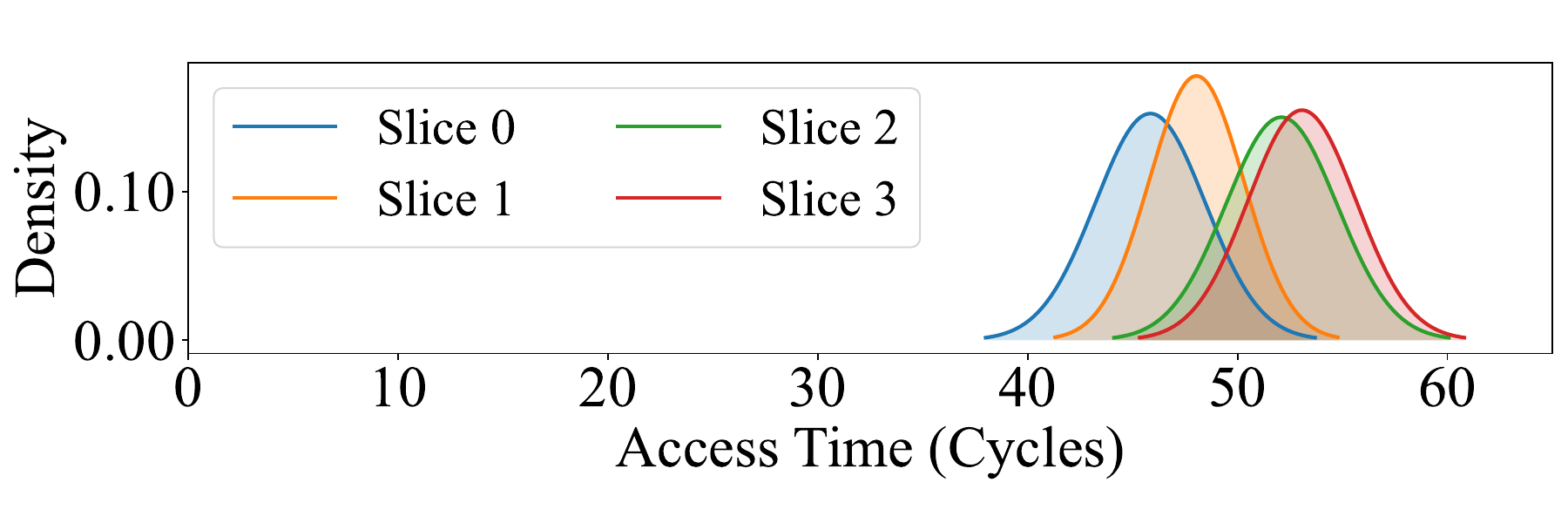}
\end{adjustbox}
\end{center}
\caption{LLC slice access latency probability distributions for the four slice i7-6700K, measured from core zero.}
\label{fig:llc-slice-timings}
\end{figure}

\subsubsection{LLC Slice Access Time Variance}
To start our investigation, we first validate that LLC slices indeed exhibit different access times.
We execute the experiment on an Intel i7-6700K processor, which has four physical cores and four slices.
We first create L2 eviction sets, which allow us to evict the tested memory lines from the L2 cache. 
We then set the process affinity to core 0 and use the RDTSC instruction to measure access latency from that core to the memory lines.

To determine the slice each memory line maps to, we use the \texttt{/proc/<pid>/pagemap} interface.
We then use the published LLC slice function for the i7-6700K~\cite{mauriceReverseEngineeringIntel2015} to determine ground-truth slice mappings and check the accuracy of our predictions.
In \cref{subsec:calibration-without-ground-truth} we show how to determine the slice in an unprivileged scenario, i.e.\ without access to \texttt{/proc/<pid>/pagemap}.

We average 10,000 accesses to each slice over 100 independent executions to establish their access time distributions.
We take 100 overall samples for this experiment as we observed slice timings to differ between runs due to frequency changes in the core and uncore.
This gives an indication of the global access time distribution for each slice.
We repeat individual address measurement when the reported cycles are greater than 100, indicating a RAM access due to spurious microarchitectural noise (e.g.\ cache eviction due to system interrupts).

\cref{fig:llc-slice-timings} shows the distribution of access times to each slice. 
We note that the access times distributions differ between slices, although they overlap.
This implies that information from access times can be used to predict the slice index of a memory address.

\begin{figure}[!t]
\begin{center}
\begin{adjustbox}{max height=4.5cm}
\includegraphics{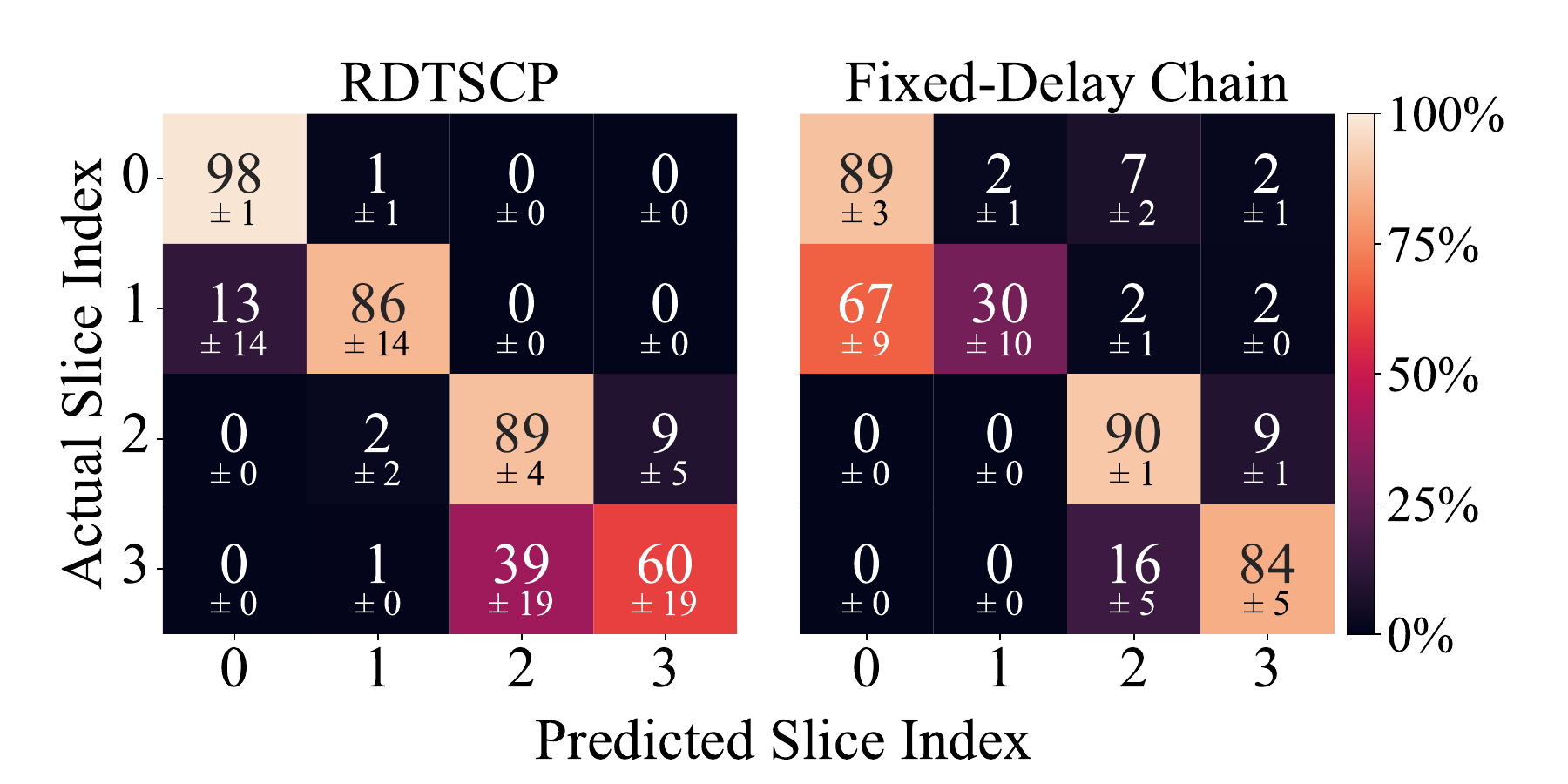}
\end{adjustbox}
\end{center}
\caption{RDTSCP and fixed-delay chain slice classification accuracy in a low-noise system, measured from core zero.}
\label{fig:llc_rdtscp_NOT_gate_quiet_predictions}
\end{figure}

\subsubsection{RDTSCP Predictions}
We now reverse the scenario.
Instead of trying to find the time to access a slice, we attempt to determine the slice by measuring the access time.
To calibrate our experiment, we first compute the average access time to each slice, averaging over 1,000 measurement for each slice.
We then measure the access time to 10,000 memory addresses, and compare the access time to the calibrated measurements to predict the slice each memory address maps to.
Due to the overlap observed in \cref{fig:llc-slice-timings}, a single timing measurement is likely to be a bad predictor.
To reduce the noise, we average the result of 10 measurements and use the average to predict the slice.
As before, we ignore  outliers with a value above 100 cycles.

The confusion matrix in the left side of \cref{fig:llc_rdtscp_NOT_gate_quiet_predictions} shows  the slice prediction accuracy and the standard deviations.
We achieve an average prediction true positive accuracy of 83\% by normalising the confusion matrix to percentages.
While averaging timing samples enhances accuracy, there is still room for improvement, as it is not optimal for all slices.
Notably, this method cannot easily differentiate between slices two and three due to their overlapping access times.
Although the attacker could use more advanced timing analysis techniques to improve accuracy~\cite{10.1145/1455526.1455530, 9152782, cryptoeprint:2023/1441}, these require in-depth analysis of the noise distribution.

\subsubsection{Weird Gates for Access Timing}\label{subsec:weird-gates-for-access-timing}
We now propose a different access time measurement technique using recent weird gate concepts.
Recall that such gates use microarchitectural races to enact certain behaviour such as logical operations, re-purposing the memory hierarchy to perform computation.
We seek to exploit this behaviour to instead measure slice access times with higher resolution than RDTSCP\@.

Our core idea is that the NOT gate of \citet{kaplanOptimizationAmplificationCache2023}, illustrated in \cref{fig:not-gate}, already compares the execution time of a memory access to that of a fixed delay.
To that aim, we place \texttt{input} in the LLC, but evict it from the L1 and the L2 caches.
Our aim is to rely on the difference in access time to various slices in order to determine in which LLC slice \texttt{input} is cached. 
For that, we try to set the length of the delay in the gate such that the access to \texttt{output} loses the race if \texttt{input} is cached in a closer LLC slice, but wins it if \texttt{input} is in a further slice.

\begin{figure}[!t]
\begin{center}
\begin{adjustbox}{width=\linewidth}
\includegraphics{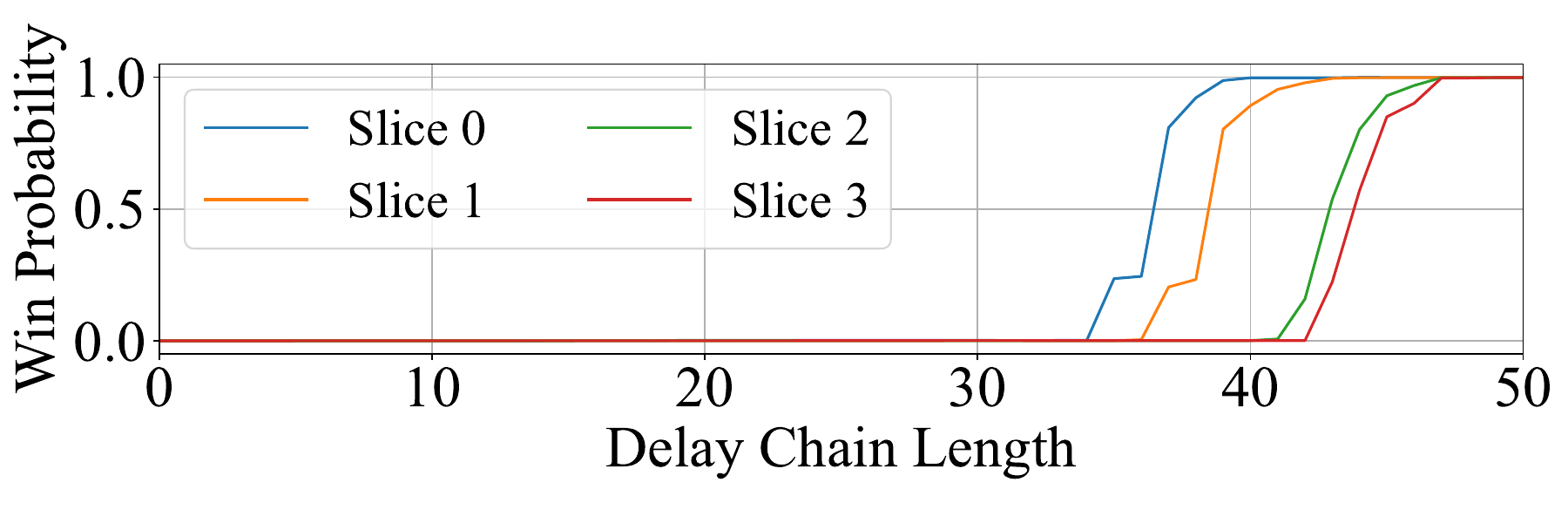}
\end{adjustbox}
\end{center}
\caption{Fixed-delay chain gate win probability as a function of the delay chain length, measured from core zero.}
\label{fig:NOT-gate-probabilities}
\end{figure}

We construct the delay from a chain of dependent \texttt{add} instruction, which we assume take one cycle each.
Increasing the number of \texttt{add} instructions increases the delay before the access to \texttt{output}.
We then test the gate with different \texttt{input} addresses, each residing in a different LLC slice.
\cref{fig:NOT-gate-probabilities} shows the probabilities of the delay chain winning as functions of the length of the chain.
We observe a clear distinction in tipping points for each slice, indicating that the gate can distinguish each of the four slices.
As a result, we can now try to use the delay chain length as a unit of measurement for the memory's LLC access time.

\subsubsection{Fixed-Delay Chain Predictions}
Having found that different slices show tipping points at different delay lengths, we now design an experiment to assess the possibility of using modified NOT gates to predict the slice a memory location maps to.
The core idea is to measure the tipping point for a memory location and compare it to the known tipping points for each slice.
In more details, we first calibrate the experiment, determining the tipping point for each slice.
For that, we select a memory location for which we know the slice, and search for the delay length at which the win probability of the gate changes.
We repeat the process for each slice, finding the tipping point for each.

We then attempt to predict the slices of 10,000 memory addresses.
For each address, we search for the tipping point. Once found, we compare to the tipping points obtained during the calibration and select the closest one as a prediction for the slice.
\cref{fig:llc_rdtscp_NOT_gate_quiet_predictions} (right) shows the prediction success.
Although this method can distinguish between slices two and three better than RDTSCP, its average true positive accuracy is not as high as overall at 73\%.
In particular, it appears that this method fails to accurately distinguish between slices one and two.

\begin{figure}[!t]
\begin{center}
\begin{adjustbox}{max height=4.5cm}
\includegraphics{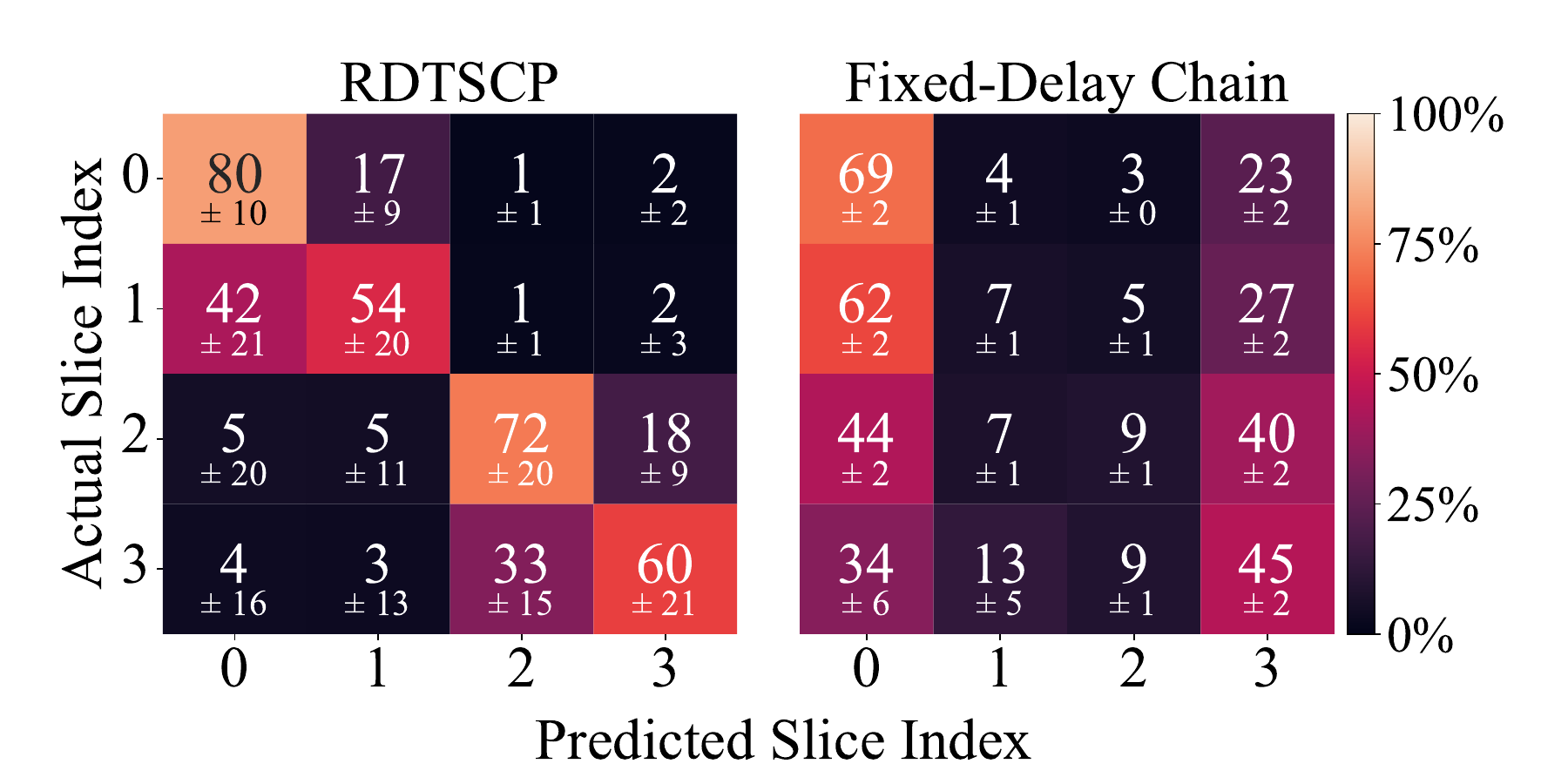}
\end{adjustbox}
\end{center}
\caption{RDTSCP and fixed-delay chain slice classification accuracy in a busy system, measured from core zero.}
\label{fig:llc_rdtscp_NOT_gate_busy_predictions}
\end{figure}

\subsection{Slice Prediction Under Noise}\label{subsec:when-these-methods-fail}
The experiments in \cref{subsec:llc-slice-access-latency} were undertaken under relatively quite system conditions, with no significant system activity.
To evaluate slice prediction in more realistic scenarios, we now repeat the experiments in the presence of activity.
We employ \textit{stress-ng}~\cite{kingColinIanKingStressng2024} to generate uncore traffic with a cache stress preset running on all non-measuring cores.
All elements of either experiment remain the same, except for the increase in system activity.

\cref{fig:llc_rdtscp_NOT_gate_busy_predictions} summarises the results.
For RDTSCP, the prediction accuracy now drops from the average of 83\% in the quiet scenario to 67\% in the noisy scenario.
The granularity of this timer is too coarse, with overlapping predictions for slices 0--1 and 2--3.
The variance in access times likewise increases, causing prediction errors.
The prediction accuracy of the weird gate method drops even further, i.e. from 73\% in the quiet scenario
to only 32\% in the noisy scenario, as shown in the right side of \cref{fig:llc_rdtscp_NOT_gate_busy_predictions},
indicating that this measurement technique is overly sensitive to the processor's frequency and noise.
In conclusion, although the techniques we investigate in  \cref{subsec:llc-slice-access-latency} have an acceptable accuracy when the system is idle, their accuracy drops significantly when the system is busy.

\begin{figure}[!t]

\footnotesize
\setlength{\ygrid}{2cm}
\setlength{\xgrid}{2.2cm}
\begin{center}
\adjustbox{width=0.85\linewidth}{
\begin{tikzpicture}[-stealth', scale=0.7]
  \tikzstyle{textbox} = [rectangle, minimum height=.5cm,text width=0.5\xgrid, minimum width=0.5\xgrid, text centered, draw=black, rounded corners =.1cm]
  \tikzstyle{chain} = [textbox, rounded corners=0cm]
  \tikzstyle{nospec} = [fill=lightgray!50!white]
  \node(FENCE){Start};
  \node[textbox] at (0,-\ygrid) (IN){Access \texttt{input}};
  \node[textbox, chain, anchor=south] at (0,-2.4\ygrid) (RETADR) {\footnotesize Set return address};
  \node[minimum height=0.25\ygrid, minimum width=0.6\xgrid, nospec, anchor=south] at (\xgrid, -2.77\ygrid) {};
  \draw[-, dotted, line width=.5mm] (-.9\xgrid, -2.4\ygrid) -- (1.7\xgrid, -2.4\ygrid);
  \node[anchor=north west, text width=1\xgrid] at (-\xgrid, -2.4\ygrid) {\footnotesize Misprediction\\detected};
  \draw (FENCE) --coordinate[midway](S1P) (IN);
  \draw (IN)-- (RETADR);
  \node[textbox, minimum height=1.4\ygrid] at (1\xgrid,-1.75\ygrid) (STR1){Access \texttt{compare}};
  \draw[dashed] (S1P) -|  (STR1);
  \node[chain, nospec, minimum height=0.6\ygrid, anchor=south] at (1\xgrid, -3.9\ygrid)(DELAY1) {Delay};
  \draw[dashed] (STR1) -- (DELAY1);
  \node[textbox, nospec] at (1\xgrid,-4.45\ygrid) (SIG1) {Access \texttt{signal}};
  \draw[dashed] (DELAY1) -- (SIG1);

  \begin{scope}[xshift=3\xgrid]
    \node(FENCE){Start};
    \node[textbox, minimum height=1.4\ygrid] at (0,-1.75\ygrid) (IN){Access \texttt{input}};
    \node[textbox, chain, anchor=south] at (0,-3.8\ygrid) (RETADR) {\footnotesize Set return address};
    \draw[-, dotted, line width=.5mm] (-.9\xgrid, -3.8\ygrid) -- (1.7\xgrid, -3.8\ygrid);
    \node[anchor=north west, text width=1\xgrid] at (-\xgrid, -3.8\ygrid) {\footnotesize Misprediction\\detected};
    \draw (FENCE) --coordinate[midway](S1P) (IN);
    \draw (IN)-- (RETADR);
    \node[textbox] at (1\xgrid,-1.0\ygrid) (STR1) {Access \texttt{compare}};
    \draw[dashed] (S1P) -| (STR1);
    \node[chain, minimum height=0.6\ygrid, anchor=south] at (1\xgrid, -2.4\ygrid)(DELAY1) {Delay};
    \draw[dashed] (STR1) --  (DELAY1);
    \node[textbox] at (1\xgrid,-2.95\ygrid) (SIG1) {Access \texttt{signal}};
    \draw[dashed] (DELAY1) -- (SIG1);

  \end{scope}
\end{tikzpicture}
}
\caption{Operation of the comparator weird gate. Left when \texttt{input}'s latency is less than \texttt{compare}, right when it is not. Shaded instructions never execute.}
\label{fig:comparator-gate}
\end{center}
\end{figure}
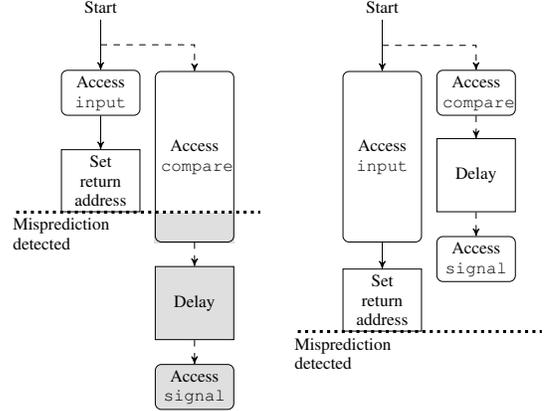

\subsection{The Comparator Weird Gate}\label{subsec:the-comparator-weird-gate}
We now present the comparator gate, a new weird gate design that allows predicting the LLC slice of a memory address even in the presence of system noise.
We observe the underlying cause of the noise is that the access latency to slices is not fixed but depends on environmental factors, such as fluctuating core frequencies and varying uncore load.
In contrast, both approaches presented in \cref{subsec:llc-slice-access-latency} do not account well to these factors.
The RDTSCP approach measures latency against the wall clock, which has a fixed frequency, whereas the  weird gate measures latency against the core clock, which is independent of the uncore clock and load.
Because variations due to environmental factors are bigger than the differences between access latencies to different LLC slices, measurements that do not account for environmental factors are doomed to fail.

To overcome this challenge, we propose to avoid independent clocks and instead compare the access time to a memory address against the instantaneous access latency to the LLC slices.
For that purpose, we design the \emph{comparator} weird gate, demonstrated in \cref{fig:comparator-gate}.
Unlike prior weird gates~\cite{katzmanGatesTimeImproving2023, WangPWB24, kaplanOptimizationAmplificationCache2023}, which compare a memory access to a fixed-delay chain, our comparator gate races two independent memory accesses.
Specifically, our comparator gate creates a microarchitectural race between the access times to two memory addresses: an \texttt{input} address, which we wish to determine the slice it maps to, and a \texttt{compare} address, which is in a known slice.
We use a \texttt{signal} to indicate whether the access to \texttt{compare} takes longer than the access to \texttt{input}.
To compensate  for the time required for setting up the return address, we need to add a small delay after accessing \texttt{compare}.
We experimentally determine the length of this delay to balance the lengths of the chains when accessing addresses in the same slice.
We provide the implementation of the gate in \cref{app:comparator-gate}.

\begin{figure}[!t]
\begin{center}
\begin{adjustbox}{width=\linewidth}
\includegraphics{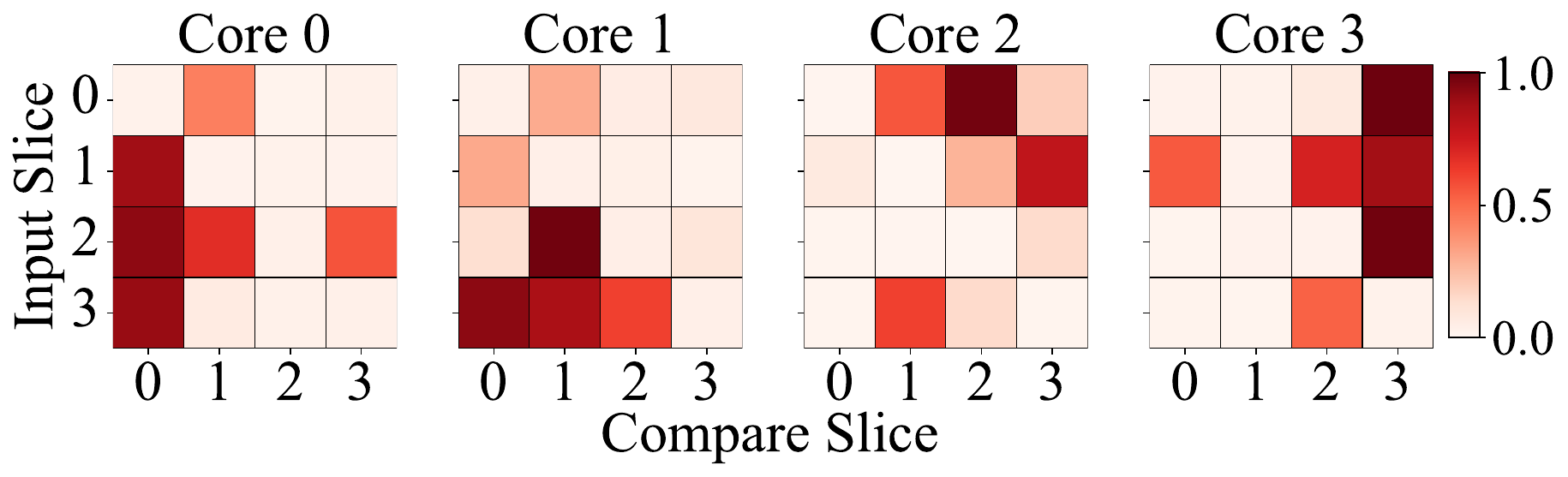}
\end{adjustbox}
\end{center}
\caption{Average comparator gate \texttt{signal} access probability when compared across each slice and core. Darker indicates the signal memory was accessed, and hence \texttt{compare} slice access was faster.
}
\label{fig:slice-comparator-gate-timings}
\end{figure}

To determine whether the comparator gate can distinguish between different slices, we test all combinations of \texttt{input} and \texttt{compare} slices.
For the experiment, we use 10,000 addresses, taking 10 measurements for each address from each core.
\cref{fig:slice-comparator-gate-timings} shows the average signal value.
As the figure demonstrates, the patterns of the signal values for each \texttt{input} slice depend on the core the experiment runs on.
However, on each core different \texttt{input} slices show different signal patterns.
Thus, we can use signal patterns as predictor for the slice.

\subsection{Comparator Gate Evaluation}\label{subsec:comparator-gate-evaluation}
We now evaluate whether the comparator gate is suitable for slice predictions.
For that, we use a set of compare addresses, one for each slice.
In this section, we rely on known ground-truth for generating this set.
\cref{subsec:calibration-without-ground-truth} shows how to generate the compare set without elevated privileges. 

As \cref{fig:slice-comparator-gate-timings} shows, an input address does not always win or always lose the race against a given compare address.
Instead, there is a win probability that depends on the slices of both the input and the compare addresses.
Consequently, to determine the slice an input address maps to, we need to execute the comparator gate multiple times for each pair of addresses to determine the win probability.

Our method for determining the slice a memory address maps to consists of two main steps.
In the profiling step, we measure the win probability for inputs of known slices against the memory addresses in the compare set. 
The profile consists of a set of vectors, one for each input slice.
The vector includes one coordinate for each compare slice, whose value is the win probability for an input of the given input slice against the compare address in the compare slice.
To ensure the accuracy of the profile, we use 1,000 invocations of the comparator gate to determine each vector coordinate.
\cref{fig:slice-comparator-gate-timings} can be viewed as visualising four such profiles, one for each core.
Each vector in the profile is visualised as a horizontal line matching the input slice.

Once the profile is created, we move to the second step of predicting the slice of a memory address.
For that, we use the comparator gate to test the input memory address against each of the addresses in the compare set.
We repeat the measurement 10 times for each compare slice, creating a probability vector that indicates how likely is the input slice to win against each of the elements in the compare set.
We then compute the Euclidean distance between this vector and each of the vectors in the profile and use the input slice of the closest vector as the predicted slice index.

\begin{figure}[!t]
\begin{center}
\begin{adjustbox}{max height=4.5cm}
\includegraphics{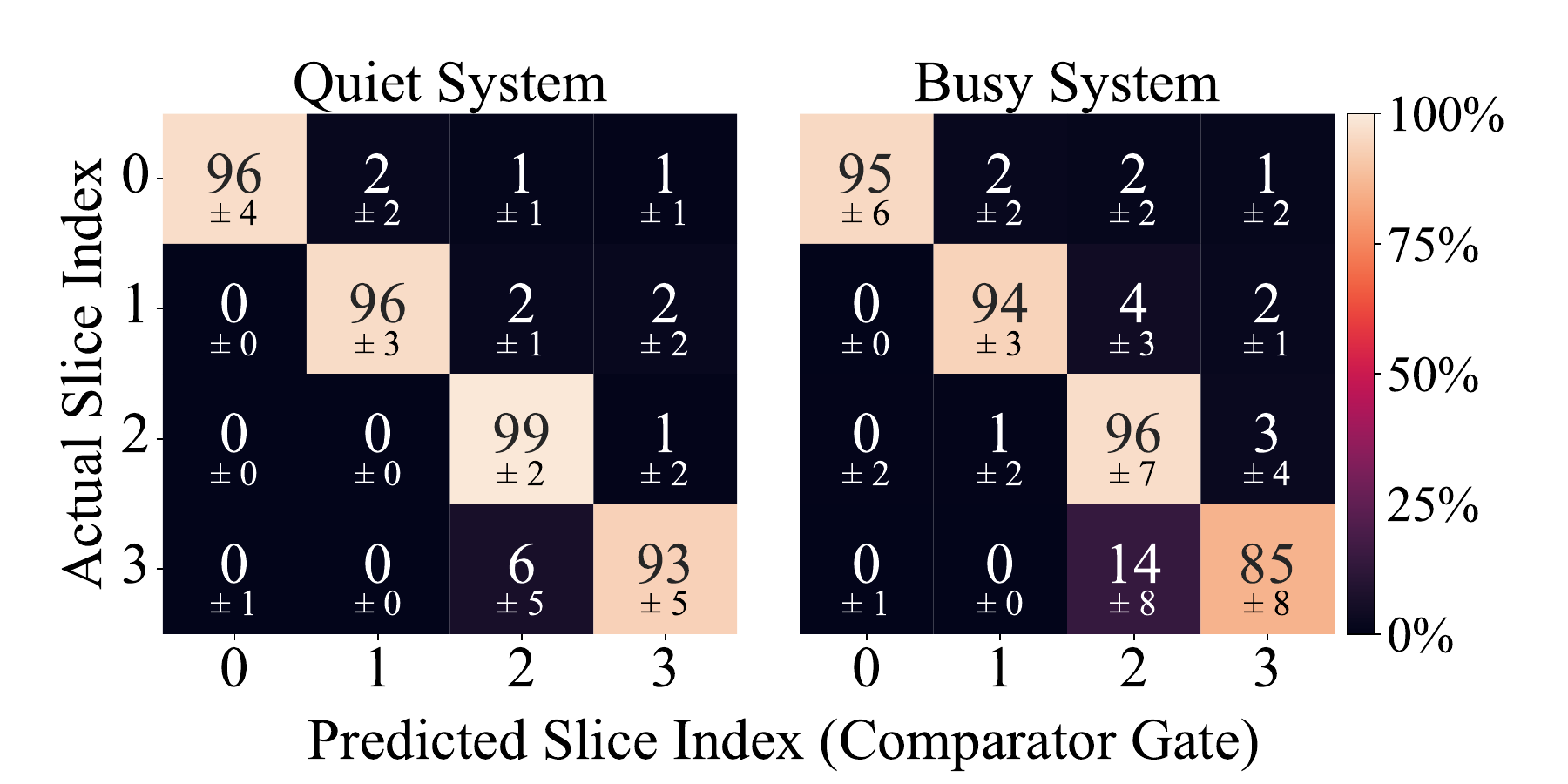}
\end{adjustbox}
\end{center}
\caption{Comparator gate slice classification accuracy, measured from core zero.}
\label{fig:slice-comparator-gate-predictions}
\end{figure}

\cref{fig:slice-comparator-gate-predictions} displays the resulting slice prediction confusion matrix.
It shows that our method does not suffer from the same noise-related accuracy loss as RDTSCP and the fixed-delay chain gate.
This gate demonstrates 96\% and 93\% overall prediction accuracy in the quiet and the busy scenarios respectively.

\parhead{Multi-core.}
As \cref{fig:slice-comparator-gate-timings} shows, the profile depends on the core on which we execute the experiment.
To evaluate our slice prediction method with the different profiles, we repeat the experiment running on each processor core.
\cref{tab:per-core-comparator-gate-prediction-percentages} summarises the results, showing the true positive classification rates for each slice.
Our method maintains a 97\% true positive rate when averaged across all cores and slices, showing that it can adapt to any processor affinity.

\begin{table}[t]
\caption{True positive comparator gate classification percentage rates for each slice and core on a quiet system.}
\centering
\begin{adjustbox}{width=\linewidth}
\begin{tabular}{r*{5}{r@{$~\pm$}l}}
\toprule
\textbf{Core} & \multicolumn{2}{c}{\textbf{Slice 0}} & \multicolumn{2}{c}{\textbf{Slice 1}} & \multicolumn{2}{c}{\textbf{Slice 2}} & \multicolumn{2}{c}{\textbf{Slice 3}} & \multicolumn{2}{c}{\textbf{Average}} \\
\midrule
0 & $      96 $ & $  4$ & $      96 $ & $  3$ & $      99 $ & $  2$ & $      93 $ & $  5$ & $      96 $ & $  4$ \\
1 & $      99 $ & $  1$ & $      95 $ & $  5$ & $      95 $ & $  2$ & $      99 $ & $  5$ & $      97 $ & $  3$ \\
2 & $      98 $ & $  6$ & $      95 $ & $  3$ & $      94 $ & $  6$ & $      99 $ & $  1$ & $      97 $ & $  4$ \\
3 & $      93 $ & $  7$ & $      98 $ & $  4$ & $      95 $ & $  8$ & $      96 $ & $  8$ & $      96 $ & $  7$ \\
\bottomrule
\end{tabular}
\end{adjustbox}
\label{tab:per-core-comparator-gate-prediction-percentages}
\end{table}

\parhead{Profile Stability.}
The attacker is unlikely to have an a-priory knowledge of the core they execute on.
Moreover, we find that the profile may vary between executions, even on the same core.
To allow executing the attack without knowledge of the core and to overcome profile variations,
 we compute the profile before each execution of our eviction-set construction algorithm.

\section{Intra-Page Slice Mappings Propagation}\label{sec:slice-decision-tree}
In \cref{sec:transient-address-access-timing}, we discuss our method for classifying memory by slice.
This requires measuring every address in a page, which is time-consuming while having a non-zero error rate.
In this section, we leverage knowledge of the slice function's structure to assist in two key areas.
We first introduce our technique for determining our compare set addresses for the comparator gate to run our profiling step.
Then, we describe how we can propagate slice classifications within a memory page to reduce the number of measurements required.

We observe that for each virtual memory page, the function implies a finite number of mappings between page offsets and LLC slices, which we refer to as a page-slice mapping.
These mappings result from the dependency of the slice function on both the known lower  physical address bits (the offset in the page) and the unknown upper address bits.
Our goal is to determine the correct page-slice mapping for a given page.
Our first method, \emph{closest match}, involves using our comparator gate to map all offsets in a page to their corresponding slices and selecting the mapping which matches the most measured slice indices.
The next, \emph{Bayesian inference}, determines the most likely page mapping based on a dynamic probability model measuring only a subset of offsets.
The final method, \emph{decision tree}, consists of a guided search based on maximising information gain to find the page-slice mapping with the fewest measurements.
We evaluate these methods on several Intel processors, determining the overall execution speed and resulting accuracy in comparison to the default comparator gate method.

\parhead{Structure of the Slice Function.}
Recall that the LLC slice function is a proprietary function used by Intel processors to determine the slice index for a given physical address.
The main property we want to exploit is that the slice function generates mappings of physical memory cache line offsets to slice indices.

Several prior works detail the observed structure of the proprietary function which  uses either a linear or non-linear approach to calculate the slice index depending on the number of slices and the processor~\cite{hundPracticalTimingSide2013, irazoquiSystematicReverseEngineering2015, mauriceReverseEngineeringIntel2015, inciSeriouslyGetMy2015, yaromMappingIntelLastLevel2015, mccalpinMappingAddressesL32021, gerlachEfficientGenericMicroarchitectural2023}.
For linear functions, the slice index is calculated by an XOR operation on the physical address bits using several permutation selection masks.
This equally distributes cache lines across the slices. 
In contrast, non-linear functions have a two-phase approach, starting with a linear XOR permutation selection, piped into a secondary stage which outputs the slice index in a range based on the XOR result.

We observe the number of unique page-slice mappings is equal to the number of slices for linear functions.
For a CPU with $n$ slices, this results in mapping consecutive cache line offsets in a page to the $n$ different slices.
This is due to the equal distribution design, through interactions between the XOR operation involving both known lower address bits and unknown upper bits.
For example, an i7-6700K generates the following four mappings (where mappings B, C, and D, are simply A XORed with 0x1, 0x2, and 0x3 respectively):
\begin{align*}
\text{Mapping. A:}~\texttt{0123 0123~\ldots~2301 2301} \\
\text{Mapping. B:}~\texttt{1032 1032~\ldots~3210 3210} \\
\text{Mapping. C:}~\texttt{2301 2301~\ldots~0123 0123} \\
\text{Mapping. D:}~\texttt{3210 3210~\ldots~1032 1032}
\end{align*}
For non-linear functions, the number of unique page-slice mappings is greater than the number of slices and depends on the processor.
Processors with six and ten LLC slices have 128 unique page-slice mappings.

We detail our method for reverse-engineering the slice function in \cref{app:slice-function-retrieval-algorithms}, and provide this as an auxiliary tool in our open-source repository.\footnote{\repoUrl}

\subsection{Determining the Compare Set}\label{subsec:calibration-without-ground-truth}
Our unprivileged threat model restricts us to 4\,KB memory pages with access to only the lower 12 bits of the physical address.
Consequently, we do not know the ground-truth slice mappings.
To run the profiling step for the comparator gate, we must first determine the compare set of addresses, one in each slice, despite not knowing the ground truth.
Therefore, we need to determine the slice mappings for a single virtual page of memory to find our compare set and also addresses with known slices for calibration purposes.

Processors with linear slice functions with $n$ slices generate $n$ unique page-slice mappings, where we can choose $n$ pre-determined offsets for any page which are guaranteed to map to $n$ different slices.
The rest of the offsets are used as calibration targets to initialise the comparator gate vectors.
We note that we only know that the addresses in the compare set are in different slices, but we do not know the exact mapping (i.e. which address maps to which).

We cannot use this approach for processors with non-linear functions because we cannot assume that pre-determined $n$ offsets in a page will map to $n$ different slices.
In order to find the compare set, we try to identify the slice mapping for the page we are targeting.
Once we identify this, we can select $n$ offsets that map to the $n$ slices.
We use our comparator gate to run pairwise comparisons for all offsets in the page.
This creates a comparison pattern based on closer versus further away slices.
We then count how many times the measurement pattern from the gate matches each of the page-slice mappings.
We select the mapping with the highest number of matches.
It is important to note that this approach does not fully identify the slices, as several equivalent mappings can appear from our measurements.
However, this does not prevent the technique from working, and the profile we generate can successfully distinguish different slices and partition the memory without knowing the ground truth.
For simplicity, the rest of the discussion assumes that the determined function is correct, rather than an equivalent function.

\subsection{Closest Match}\label{subsec:closest-match}
The first approach we explore as a means to increase the classification speed while maintaining accuracy for the slice mappings predictions is the closest match.
This method involves comparing all addresses in a page to the compare set using the comparator gate, then selecting the mapping which matches the majority of predicted slice indices.
The method works for both linear and non-linear slice functions.
It results in much higher accuracy than the default comparator gate method, as it uses knowledge of the slice function to make predictions.
However, as it still requires comparing each address to the comparison set, so it does not improve the classification speed.

\subsection{Bayesian Inference}\label{subsec:bayesian-inference}
Our next approach involves the use of statistical Bayesian inference to guess the page-slice mapping.
The aim here is to maintain probabilities for each page-slice mapping, and update such probabilities based on iteratively measured slice indices to determine the most likely mapping.
At the beginning, each page starts with an equal probability of occurrence as we have no prior information.
We then predict the slice index for the first offset in the page using the comparator gate.
We update the posterior probabilities of each unique page mapping given the predicted slice index and move to the next offset to repeat.
When we reach a given threshold (0.90), we select the mapping with the highest probability as the page's slice mapping and can now proceed to the next page.
This results in a fewer number of offsets measured per page, which we discuss further in \cref{subsec:evaluation-of-propagation-methods}.



\begin{table*}[!t]
\centering
\caption{Comparison of intra-page propagation methods for slice predictions. True positive classification accuracy listed as a percentage, speed in MB/s with respective standard deviations. Bold values indicate the best performance.}
\begin{tabularx}{\linewidth}{lr *{17}{r@{$~\pm$}X}}
\toprule
                   &        & \multicolumn{4}{c}{\textbf{Comparator Gate}} & \multicolumn{4}{c}{\textbf{Closest Match}} & \multicolumn{4}{c}{\textbf{Bayesian Inference}}     & \multicolumn{4}{c}{\textbf{Decision Tree}}  \\ \cmidrule(l){3-6} \cmidrule(l){7-10} \cmidrule(l){11-14} \cmidrule(l){15-18}
\textbf{Processor} & \textbf{Linear} & \multicolumn{2}{c}{\hspace{-1.75em}Accuracy} & \multicolumn{2}{c}{\hspace{-2em}MB/s} & \multicolumn{2}{c}{\hspace{-1.75em}Accuracy} & \multicolumn{2}{c}{\hspace{-2em}MB/s} & \multicolumn{2}{c}{\hspace{-1.75em}Accuracy} & \multicolumn{2}{c}{\hspace{-1.5em}MB/s} & \multicolumn{2}{c}{\hspace{-1.75em}Accuracy} & \multicolumn{2}{c}{\hspace{-1.75em}MB/s} \\
\midrule
i7-6700K           & \cmark &  84 &   4 &  70 &   1 & 100 &   1 &  70 &   1 &  \textbf{94} &   \textbf{4} & \textbf{568} &  \textbf{46} &  90 &   7 & 538 &  38 \\
i7-11700KF         & \cmark &  77 &  11 &  27 &   0 &  97 &   4 &  27 &   0 &  \textbf{79} &  \textbf{10} & \textbf{323} &  \textbf{20} &  71 &  11 & 352 &  23 \\
i7-13700H          & \cmark &  58 &   9 &  21 &   1 &  96 &   3 &  21 &   0 &  \textbf{74} &  \textbf{10} & \textbf{175} &  \textbf{31} &  52 &   9 & 198 &  54 \\
i7-8700            & \xmark &  79 &  11 &  45 &   6 &  98 &   7 &  48 &   1 &  98 &   7 &  70 &   4 &  \textbf{91} &   \textbf{9} & \textbf{119} &  \textbf{32} \\
i7-9850H           & \xmark &  79 &   9 &  51 &   1 &  68 &  10 &  50 &   0 &  99 &   6 &  72 &   4 &  \textbf{95} &   \textbf{5} & \textbf{187} &  \textbf{31} \\
i9-10900K          & \xmark &  74 &   9 &  32 &   6 &  99 &   4 &  34 &   0 &  99 &   5 &  60 &   3 &  \textbf{96} &   \textbf{3} & \textbf{ 96} &  \textbf{21} \\
i9-12900KF         & \xmark &  53 &   9 &  25 &   0 &  95 &   4 &  25 &   0 &  \textbf{93} &   \textbf{4} &  \textbf{41} &   \textbf{2} &  74 &   6 &  19 &   4 \\
i9-14900K          & \xmark &  52 &  12 &  15 &   0 &  96 &   6 &  15 &   0 &  \textbf{95} &   \textbf{7} &  \textbf{19} &   \textbf{1} &  82 &   8 &   6 &   2 \\
\bottomrule
\end{tabularx}
\label{tab:other-procs-decision-tree}
\end{table*}

\subsection{Decision Tree}\label{subsec:building-the-decision-tree}
\begin{figure}[!t]
\footnotesize
\setlength{\ygrid}{2cm}
\setlength{\xgrid}{2.2cm}
\begin{center}
\adjustbox{width=0.7\linewidth}{
\begin{tikzpicture}[-stealth', scale=0.7]
    \tikzstyle{node} = [rectangle, rounded corners, minimum width=1cm, minimum height=1cm,text centered, draw=black, fill=gray!10]
    \tikzstyle{leaf} = [rectangle, minimum width=1.5cm, minimum height=1cm, text centered, draw=black, fill=gray!30]
    \tikzstyle{arrow} = [thick,->,>=stealth]

    \node[node, label={[font=\bfseries]above:Root}] (root) {Offset 0x0};
    \node[node, below of=root, xshift=-2.5cm, yshift=-1.0cm] (node1) {Offset 0x100};
    \node[leaf, below of=node1, xshift=-2.5cm, yshift=-1.0cm] (leaf1) {Page Perm. A};
    \node[leaf, below of=node1, yshift=-1.5cm] (leaf2) {Page Perm. B};
    \node[leaf, below of=node1, xshift=2.5cm, yshift=-1.0cm] (leaf4) {Page Perm. C};
    \node[leaf, below of=root, xshift=1.0cm, yshift=-1.0cm] (leaf3) {Page Perm. D};

    \draw[arrow] (root) -- (node1) node[midway, left, text width=1.2cm, xshift=-0.1cm] {Predicted Slice: 0};
    \draw[arrow] (root) -- (leaf3) node[midway, right, text width=1.2cm, xshift=0.3cm] {Predicted Slice: 5};
    \draw[arrow] (node1) -- (leaf1) node[midway, left, text width=1.2cm, xshift=-0.2cm] {Predicted Slice: 1};
    \draw[arrow] (node1) -- (leaf2) node[midway, right, text width=1.2cm, xshift=0cm] {Predicted Slice: 2};
    \draw[arrow] (node1) -- (leaf4) node[midway, right, text width=1.2cm, xshift=0.3cm] {Predicted Slice: 4};
\end{tikzpicture}
}
\caption{Example decision tree structure. Nodes represent page offsets to predict the slice for. Leaves represent the determined page-slice mapping.}
\label{fig:decision_tree}
\end{center}
\end{figure}
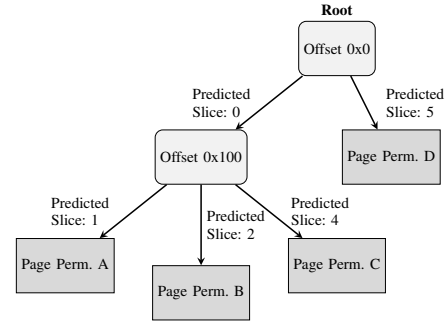

The decision tree method represents a guided search of slice indices through specific page offsets to determine the overall page-slice mapping.\footnote{This method is equivalent to the ID3 algorithm by \citet{10.1007/BF00116251}.}
We illustrate this structure in \cref{fig:decision_tree}.
Its design maximises the information gained at each decision node, reducing the number of overall slice predictions.

\parhead{Building the Tree.}
Generating the decision tree involves a recursive construction of each decision branch as we narrow down to each unique page-slice mapping.
To construct the first node, we select the page offset which provides the highest distinguishing entropy across all possible mappings. Note that for linear mapping functions, all offsets have the same distinguishing entropy at the beginning.
However, this is not necessarily the case for non-linear mappings.

Specifically, to calculate the entropy for an offset \textit{o}, we count the occurrence of every slice index at the offset for each of the \textit{n} mappings \( P \):
\begin{align*}
    \forall i \in \{0, 1 \ldots, n-1\}, \quad v &= P_{i,o} \\
    \mathit{slice\_counts}_{v} \mathrel{+}&= 1
\end{align*}

Then by using Shannon's formula, the offset's uncertainty \textit{H} is calculated by:
\[
H = - \sum_{s=0}^{\mathit{LLC\_SLICES}-1} p(s) \log_{2}p(s)
\]

where \( p(s) \) is the probability of the slice index occurring:
\[
p(s) = \frac{\mathit{slice\_counts}_s}{n}
\]

From this, we can select the offset with the highest entropy.
We then use our gate to compare this offset to our comparison set.
Then, we create new child nodes for each possible slice index, forming the edges of the tree, and remove the offset from the list of potential measurements.
With each child node, we repeat the above process using the subset of page-slice mappings which matches the slice index associated with that node.
The recursion continues until only one mapping remains.
At this point, we create a leaf node to store the unique mapping, which we use to determine the slice function for that page.

The user can select multiple offsets to measure at each decision node, which can improve the accuracy of the method at the detriment of speed.
In practice, we find that the number of offsets to measure per decision node depends on the processor and the slice function.
We provide configurations for several processors in our codebase.

\parhead{Usage.}
Starting from the root node, we compare the indicated offset to the comparison set and predict the slice.
We then traverse to the child node corresponding to the predicted slice and repeat until we reach a leaf node with a page-slice mapping.
If a child for the predicted slice does not exist, this constitutes a slice prediction error.
In this case, we traverse up to the parent node and retry.
However, this may still cause the wrong mapping to be selected if an uncaught error occurs higher up in the tree.
We leave this for future improvement.

\subsection{Evaluation of Propagation Methods}\label{subsec:evaluation-of-propagation-methods}
Using the same methodology as in \cref{sec:transient-address-access-timing}, we evaluate the propagation methods for their accuracy and classification speed.
We first place our desired memory into the LLC using L2 eviction sets, and use the propagation methods to determine the slice mappings for each page.
For each method, we measure 10,000 cache lines across the LLC, and average the results over 100 runs to get a global view of their performance.
As the decision tree can be built statically, we do not record the time taken to generate the tree in these measurements.

We compare the classification speed and accuracy of each method across several Intel processors, summarising the results in \cref{tab:other-procs-decision-tree}.
The best improvement in accuracy comes from the closest match method, which is to be expected considering that the method combines the most measurements with the knowledge of the slice function.

For linear slice functions, the dynamic Bayesian inference method is the fastest and most accurate due to the nature of the hash function.
On our i7-6700K processor, we find this method requires measuring between two and three addresses per page on average  to determine the page mapping with confidence.
For the i7-11700K and i7-13700H processors, this increases to  five offsets per page on average.

The decision tree method significantly improves the classification speed for most processors while maintaining a good level of accuracy, working especially well for non-linear functions.
We reach above 90\% accuracy for these processors.
On the i9-12900KF and i9-14900K, the decision tree method performs relatively poorly due to poor baseline comparator gate performance.
For such processors, we choose the Bayesian inference method, which achieves higher speed and accuracy because it dynamically adjusts the number of measurements required when trying to determine the page-slice mapping.


\section{Slice-Aware Eviction Set Generation}\label{sec:slice-aware-eviction-set-creation}
To successfully undertake a contention-based cross-core attack, the attacker must first generate eviction sets for every LLC set.
In this section, we introduce three optimisations for this procedure, leveraging our foreknowledge of memory slice mappings.

The first, \emph{candidate set slice filtering}, reduces initial eviction candidate sets to only contain memory with the same slice as the target address.
The second, \emph{non-linear eviction set propagation} unlocks the ability to mirror eviction sets to other page offsets on processors with non-linear slice functions.
The third, \emph{test eviction filtering} accelerates the search for a target address for which we need to build a new eviction set.
We demonstrate a significant improvement in execution time for processors with both linear and non-linear slice functions, building on state-of-the-art eviction set generation techniques e.g.\ L2 candidate set filtering and parallel probing~\cite{zhaoLastLevelCacheSideChannel2024} and outperforming Prune+PlumTree~\cite{kessousPrune+PlumTreeFindingEviction2024}.

\subsection{Implementation Details}\label{subsec:implementation-details}
Before describing our optimisations, we first  detail some techniques and configurations that we use in the implementation of our eviction set creation methodology.

\parhead{Test Eviction Function.}
Current eviction set creation methods use a \emph{test eviction} function to determine whether the given set of addresses can evict a target address from the LLC\@.
Similar to \citet{zhaoLastLevelCacheSideChannel2024}, we place all eviction set addresses in an array.
This lowers the execution time by exploiting memory-level parallelism, making this method more efficient and noise-resilient in comparison to a linked list approach~\cite{vilaTheoryPracticeFinding2019, purnalPrimeScopeOvercoming2021}.
We access the elements in the array with a user-configurable traversal pattern to suit the processor's replacement policy.
We also access another virtual address in the same page prior to measurement to ensure caching of the TLB entry, reducing false-positives for eviction~\cite{10.1007/978-3-319-93387-0_5}.

\parhead{Choice of Pruning Algorithm.}
We use the optimised group testing pruning algorithm from \citet{zhaoLastLevelCacheSideChannel2024}.
We find this version of the pruning algorithm to be more efficient than the original using early termination~\cite{vilaTheoryPracticeFinding2019}.
The original group testing algorithm uses an early termination approach, identifying and pruning removable groups as soon as they are found, without needing to search through all remaining groups.
We also find that the optimised version performs better  than the binary search algorithm~\cite{zhaoLastLevelCacheSideChannel2024}.

\parhead{L2 Candidate Set Filtering.}
We integrate the L2 candidate set filtering technique~\cite{zhaoLastLevelCacheSideChannel2024} into our eviction set generation process.
The approach builds on the observation that if two addresses map to the same LLC set, they also map to the same L2 set.
Consequently, filtering by L2 set reduces the number of addresses for the pruning algorithm to check for congruency with the target address.

\subsection{Our Optimisations}\label{subsec:slice-aware-optimisations}
Building on our insights  from \cref{sec:transient-address-access-timing} and \cref{sec:slice-decision-tree}, we implement three specific optimisations to speed up the generation of eviction sets for the entire LLC\@.

\parhead{Candidate Set Slice Filtering.}
We carry out slice filtering for our candidate set addresses to minimise the execution time of the pruning algorithm, similar to L2 candidate set filtering.
We use our knowledge of predicted slice mappings to filter the initial candidate set to consist of addresses that map to the same slice as the target address.
This increases the likelihood of finding congruent addresses, dividing the total size of the candidate set by the number of LLC slices.
The speedup depends on the algorithmic complexity of the chosen pruning method and the number of slices.

\parhead{Eviction Set Propagation for Non-Linear Slice Functions.}
Linear slice functions enable straightforward propagation from a single page offset to the rest of the page~\cite{liuLastLevelCacheSideChannel2015}.
For example, constructing a single eviction set starting at page offset \texttt{0x0} consists of addresses which are congruent with one another, mapping to the same LLC sub-slice and set.
If we add a fixed page offset to each of the addresses, the new \emph{mirrored} eviction set addresses will all map to the same LLC sub-slice.
This technique can be applied to any offset, meaning we only need to find eviction sets for a single page offset with linear functions.

Propagating eviction sets for non-linear slice functions in the same manner results in a faulty eviction set with addresses mapping several slices~\cite{yaromMappingIntelLastLevel2015}, being unable to cause cache contention.
To solve this problem, we first generate a single eviction set, and then generate mirrors for the rest of the page offsets.
We can use our previously obtained slice mappings (\cref{sec:slice-decision-tree}) to determine which addresses map to the same slice and check whether the mirrored eviction set maps entirely to a single slice.
This does not require further contention tests.
With this approach, we experimentally find we only need to generate on average 15\% of all LLC eviction sets for processors with ten slices, and 22\% for those with six slices using the conventional method.

\parhead{Test Eviction Filtering.}
When generating multiple eviction sets across an entire system, we need to determine if any previously found eviction sets can evict the target address to avoid duplicates.
For the i7-6700K, each single page offset has 128 potential LLC eviction sets.
This number increases to 320 for the i9-10900K due to its larger LLC\@.

Past methods tend to test every previously-generated eviction set with the potential target addresses~\cite{vilaTheoryPracticeFinding2019, songDynamicallyFindingMinimal2019}.
We optimise this process by testing the subset of eviction sets that correspond to the same slice and L2 set as the target.
This reduces the time between finding a target address and generating a new eviction set.

We can reduce the number of LLC eviction sets to test from 128 and 320 to just two in the aforementioned processors.
They each have 16 L2 eviction sets per page offset, derived from 1024 L2 sets across 64 L1 sets.
For the i7-6700K with four slices, we narrow the testing to just two eviction sets, given $\frac{128}{16 \times 4} = 2$.
For the i9-10900K with ten slices, the same property holds with $\frac{320}{16 \times 10} = 2$.

\begin{table}[t]
\caption{Comparison of different LLC cross-core attack frameworks and their prerequisites.}
\begin{tabular}{lcccc}
\hline
\textbf{Cross-Core Attack}                       & \textbf{\begin{tabular}[c]{@{}c@{}}4\,KB\\Pages\end{tabular}} & \textbf{Userspace}  & \textbf{\begin{tabular}[c]{@{}c@{}}Slice\\Aware\end{tabular}} & \textbf{\begin{tabular}[c]{@{}c@{}}No Shared\\Memory\end{tabular}}  \\ \hline
\citet{irazoquiSharedCacheAttack2015}            & \xmark                  & \xmark                  & \xmark                             & \xmark                    \\
\citet{grussCacheTemplateAttacks2015}            & \cmark                  & \cmark                  & \xmark                             & \xmark                    \\
\citet{liuLastLevelCacheSideChannel2015}         & \xmark                  & \cmark                  & \xmark                             & \cmark                    \\
\citet{yaromMappingIntelLastLevel2015}           & \xmark                  & \xmark                  & \cmark                             & \cmark                    \\
\citet{vilaTheoryPracticeFinding2019}            & \cmark                  & \cmark                  & \xmark                             & \cmark                    \\
\citet{purnalPrimeScopeOvercoming2021}           & \cmark                  & \cmark                  & \xmark                             & \cmark                    \\
\citet{10.1007/978-3-030-80825-9_14}             & \cmark                  & \cmark                  & \xmark                             & \xmark                    \\
P+PT~\cite{kessousPrune+PlumTreeFindingEviction2024} & \cmark                  & \cmark                  & \xmark                             & \cmark                    \\
L2CS~\cite{zhaoLastLevelCacheSideChannel2024}        & \cmark                  & \cmark                  & \xmark                             & \cmark                    \\
\textbf{Ours}                                    & \cmark                  & \cmark                  & \cmark                             & \cmark                    \\ \hline
\end{tabular}
\label{tab:attack-compare}
\end{table}

\begin{table*}[!t]
\centering
\caption{Comparison of prior eviction set generation methods, L2 candidate set filtering  (L2CS)~\cite{zhaoLastLevelCacheSideChannel2024} and Prime+PruneTree (P+PT)~\cite{kessousPrune+PlumTreeFindingEviction2024} on several Intel processors, averaged over 1,000 independent runs. Bold values indicate the best performance.}
\begin{tabularx}{\linewidth}{l r *{6}{r@{$~\pm$}X}}
\toprule
                     &                 & \multicolumn{6}{c}{\textbf{Page Offset (ms)}}                                                                                                                                                             & \multicolumn{6}{c}{\textbf{Full LLC (ms)}}                                                                     \\ \cmidrule(l){3-8} \cmidrule(l){9-14}
\textbf{Processor}   & \textbf{Cores}  & \multicolumn{2}{c}{\hspace{-1.75em}\textbf{L2CS}} & \multicolumn{2}{c}{\hspace{-1.75em}\textbf{P+PT}} & \multicolumn{2}{c}{\hspace{-1.75em}\textbf{Slice+Slice Baby}} & \multicolumn{2}{c}{\hspace{-1em}\textbf{L2CS}} & \multicolumn{2}{c}{\hspace{-1em}\textbf{P+PT}} & \multicolumn{2}{c}{\hspace{-1em}\textbf{Slice+Slice Baby}} \\
\midrule

i7-6700K & 4         & $     113 $ & $        9$ & $      83 $          & $       25$ & $      \mathbf{77} $ & $        \mathbf{7}$ & $     115 $ & $        7$ & $      83 $ & $       25$ & $      \mathbf{79} $ & $        \mathbf{9}$   \\
i7-11700KF & 8       & $     447 $ & $       95$ & $     475 $          & $       42$ & $     \mathbf{301} $ & $       \mathbf{85}$ & $     466 $ & $      121$ & $     475 $ & $       42$ & $     \mathbf{320} $ & $      \mathbf{102}$   \\
i7-8700 & 6          & $     190 $ & $       17$ & $     \mathbf{173} $ & $       \mathbf{71}$ & $     323 $ & $       55$ & $    6683 $ & $      443$ & $   11506 $ & $      935$ & $     \mathbf{812} $ & $      \mathbf{230}$   \\
i7-9850H & 6         & $     183 $ & $       21$ & $     \mathbf{145} $ & $       \mathbf{60}$ & $     255 $ & $       62$ & $    6302 $ & $      599$ & $    9616 $ & $      734$ & $     \mathbf{729} $ & $      \mathbf{280}$   \\
i9-10900K & 10       & $     544 $ & $      109$ & $     \mathbf{300} $ & $      \mathbf{106}$ & $     759 $ & $      170$ & $   15485 $ & $      846$ & $   19864 $ & $     1073$ & $    \mathbf{1569} $ & $      \mathbf{354}$   \\

\bottomrule
\end{tabularx}
\label{tab:results}
\end{table*}

\subsection{Evaluation and Discussion}\label{subsec:evaluation-and-discussion}
We now  evaluate our approach to generating eviction sets across a variety of Intel Core processors.
\cref{tab:attack-compare} compares our approach to other recent works in the field of cross-core attacks.
Using our slice-aware optimisations, we greatly reduce the execution time for the eviction set initialisation phase.

We perform eviction set generation across several Intel processors.
We do not significantly alter the operating state of the processors, aside from differing capacity and speeds of memory.
In each case, we initialise a candidate set buffer $3\times$ the size of the LLC\@ to ensure an even comparison across each methodology.
All hardware prefetchers are enabled.
We include the time taken to generate L2 eviction sets.

We evaluate two test scenarios, the first \emph{Page Offset} where we generate all possible LLC eviction sets only for a single page offset and the second \emph{Full LLC}, where we generate eviction sets for the entire LLC\@.
The \emph{Page Offset} scenario allows for comparing between processors regardless of slice function, representing the performance of our method without propagation to any remaining page offsets.
On the other hand, \emph{Full LLC} demonstrates the increased feasibility our optimisations provide for the execution time of the entire LLC eviction set generation.

We compare our approach to two recent works, one using L2 candidate set filtering (L2CS)~\cite{zhaoLastLevelCacheSideChannel2024}, and the other using the Prune+PlumTree (P+PT) algorithm~\cite{kessousPrune+PlumTreeFindingEviction2024}.
To ensure a fair comparison for processors with linear slice functions, we always carry out eviction set propagation as this is a previously understood optimisation~\cite{yaromMappingIntelLastLevel2015, kessousPrune+PlumTreeFindingEviction2024}.

We reimplement L2CS to cater for Intel Core processors as well as incorporate our slice prediction methodology.
The public implementation of the P-PT technique~\cite{kessousPrunePlumTreeFindingEvictionSetsatScale2023} cannot work across multiple page offset.
To counteract this, we instead build each offset individually, recording the execution time for just this portion of the program, summing them afterwards to permit comparison.

\parhead{Results.}
\cref{tab:results} presents the execution times for the various eviction set generation techniques across multiple Intel processors.

The main benefits of using our slice-aware optimisations become clear when considering the \emph{Full LLC} scenario for non-linear sliced processors.
Here, we achieve a \results{10900} speedup for the ten-slice i9-10900K, reducing the full LLC eviction set generation time from 15 and 20 seconds, with L2CS and P+PT, respectively, to \resultsSeconds{10900} seconds on average.
Likewise, we achieve \results{8700} and \results{9850} speedups for the i7-8700 and i7-9850H both with six slices.
Our approach does not outperform the other methods for \emph{Page Offset} due to the overhead of the slice classification routine.
However, for such processors with non-linear slice functions, we must find eviction sets at all page offsets, nullifying the disadvantage~\cite{yaromMappingIntelLastLevel2015}.

Considering linear slice function processors, there is little difference between \emph{Page Offset} and \emph{Full LLC} for all works, as eviction set propagation takes a negligible amount of time, mostly hidden by noise.
For the i7-6700K, our algorithm provides an improvement of \results{6700} over P+PT and 1.5$\times$ over L2CS\@.
For the i7-11700K, we achieve a \results{11700} speedup over L2CS and P+PT\@.

\parhead{Quality of Eviction Sets.}
To evaluate the quality of the eviction sets, we record the number of found eviction sets, occurrence of duplicates, and those which were missing.
A duplicate maps to the same set and slice as another eviction set.
A missing eviction set occurs when there were LLC sets which could not be evicted by any of the generated eviction sets.

In all runs, both our method and L2CS find at least 99\% of the total eviction set count for the LLC\@.
There was a negligible rate of duplicate and missing eviction sets, with ours marginally higher due to duplication of errors during eviction set propagation.

We also augment P+PT to monitor these metrics.
We find that on some processors the algorithm can experience a higher rate of missing and duplicate eviction sets.
At best, the algorithm missed 1\% of the total eviction sets for the i7-6700K, also experiencing the highest rate of duplicates at 5\%.
At worst, 14\% of the total eviction sets were missing for the i7-1100K\@.
It may be possible that further tuning of the P+PT algorithm to the processors we use would improve the results.  
We leave testing this to future work.

\section{Conclusion}\label{sec:conclusion}
We detail several enhancements to contention-based side-channel attacks by accelerating the initial generation of LLC eviction sets.
Our work removes the uncertainty posed by the proprietary LLC slice function implemented in all major Intel processor families, which otherwise prolongs the running time of the eviction set generation procedure.

We design a technique for memory slice classification that uses weird gates, allowing us to determine slice mappings in unprivileged scenarios.
Using knowledge of the reverse-engineered slice functions, we evaluate various methods for intra-page propagation of slice mappings, significantly reducing the time required to classify memory with our weird gate.

Finally, we present three optimisations for generating eviction sets for the entire LLC, leveraging knowledge of slice mappings to achieve a significant speedup in execution time over state-of-the-art.

\section*{Acknowledgements}
This work was supported by
the Air Force Office of Scientific Research (AFOSR) under award number FA9550-24-1-0079;
the Alfred P. Sloan Research Fellowship;
an ARC Discovery Project number DP210102670;
the Defense Advanced Research Projects Agency (DARPA) under contract numbers W912CG-23-C-0022,
Defence Science and Technology Group (DSTG), Australia under Agreement No.~11965;
the Deutsche Forschungsgemeinschaft (DFG, German Research Foundation) under Germany's Excellence Strategy - EXC 2092 CASA - 390781972;
ISF grant no. 1807/23;
Len Blavatnik and the Blavatnik Family Foundation;
Stellar Development Foundation;
and gifts from Cisco and Qualcomm.

The views and conclusions contained in this document are those of the authors and should not be interpreted as representing the official policies, either expressed or implied, of the U.S. Government.

\bibliographystyle{IEEEtranSN}
\bibliography{ref}

\begin{thebibliography}{64}
\providecommand{\natexlab}[1]{#1}
\providecommand{\url}[1]{#1}
\csname url@samestyle\endcsname
\providecommand{\newblock}{\relax}
\providecommand{\bibinfo}[2]{#2}
\providecommand{\BIBentrySTDinterwordspacing}{\spaceskip=0pt\relax}
\providecommand{\BIBentryALTinterwordstretchfactor}{4}
\providecommand{\BIBentryALTinterwordspacing}{\spaceskip=\fontdimen2\font plus
\BIBentryALTinterwordstretchfactor\fontdimen3\font minus
  \fontdimen4\font\relax}
\providecommand{\BIBforeignlanguage}[2]{{%
\expandafter\ifx\csname l@#1\endcsname\relax
\typeout{** WARNING: IEEEtranSN.bst: No hyphenation pattern has been}%
\typeout{** loaded for the language `#1'. Using the pattern for}%
\typeout{** the default language instead.}%
\else
\language=\csname l@#1\endcsname
\fi
#2}}
\providecommand{\BIBdecl}{\relax}
\BIBdecl

\bibitem[Brumley and Hakala(2009)]{10.1007/978-3-642-10366-7_39}
B.~B. Brumley and R.~M. Hakala, ``Cache-timing template attacks,'' in
  \emph{AsiaCrypt}, 2009.

\bibitem[Chuengsatiansup et~al.(2022)Chuengsatiansup, Genkin, Kuepper, Wagner,
  Wu, and Yarom]{chuengsatiansup0xADE1A1DEAssemblyLine2022}
\BIBentryALTinterwordspacing
C.~Chuengsatiansup, D.~Genkin, J.~Kuepper, M.~Wagner, D.~Wu, and Y.~Yarom,
  ``{0xADE1A1DE/\allowbreak Assembly\allowbreak Line},'' 2022. [Online].
  Available: \url{https://github.com/0xADE1A1DE/AssemblyLine}
\BIBentrySTDinterwordspacing

\bibitem[Crosby et~al.(2009)Crosby, Wallach, and
  Riedi]{10.1145/1455526.1455530}
\BIBentryALTinterwordspacing
S.~A. Crosby, D.~S. Wallach, and R.~H. Riedi, ``Opportunities and limits of
  remote timing attacks,'' \emph{ACM Trans. Inf. Syst. Secur.}, 2009. [Online].
  Available: \url{https://doi.org/10.1145/1455526.1455530}
\BIBentrySTDinterwordspacing

\bibitem[Didier and Maurice(2021)]{10.1007/978-3-030-80825-9_14}
G.~Didier and C.~Maurice, ``Calibration done right: Noiseless {Flush+Flush}
  attacks,'' in \emph{DIMVA}, 2021.

\bibitem[Disselkoen et~al.(2017)Disselkoen, Kohlbrenner, Porter, and
  Tullsen]{disselkoenPrimeAbortTimerFree2017}
\BIBentryALTinterwordspacing
C.~Disselkoen, D.~Kohlbrenner, L.~Porter, and D.~Tullsen, ``Prime+{Abort}: A
  timer-free high-precision {L3} cache attack using {Intel} {TSX},'' in
  \emph{USENIX Sec.}, 2017. [Online]. Available:
  \url{https://www.usenix.org/conference/usenixsecurity17/technical-sessions/presentation/disselkoen}
\BIBentrySTDinterwordspacing

\bibitem[Doweck et~al.(2017)Doweck, Kao, Lu, Mandelblat, Rahatekar, Rappoport,
  Rotem, Yasin, and Yoaz]{doweck6thGenerationIntelCore2017}
J.~Doweck, W.-F. Kao, A.~K.-y. Lu, J.~Mandelblat, A.~Rahatekar, L.~Rappoport,
  E.~Rotem, A.~Yasin, and A.~Yoaz, ``Inside 6th-generation {Intel} {Core}: New
  microarchitecture code-named {Skylake},'' \emph{IEEE Micro}, vol.~37, no.~2,
  2017.

\bibitem[Evtyushkin et~al.(2021)Evtyushkin, Benjamin, Elwell, Eitel, Sapello,
  and Ghosh]{evtyushkinComputingTimeMicroarchitectural2021}
D.~Evtyushkin, T.~Benjamin, J.~Elwell, J.~A. Eitel, A.~Sapello, and A.~Ghosh,
  ``Computing with time: Microarchitectural weird machines,'' in \emph{ASPLOS},
  2021.

\bibitem[Genkin et~al.(2018)Genkin, Pachmanov, Tromer, and
  Yarom]{10.1007/978-3-319-93387-0_5}
\BIBentryALTinterwordspacing
D.~Genkin, L.~Pachmanov, E.~Tromer, and Y.~Yarom, ``Drive-by key-extraction
  cache attacks from portable code,'' in \emph{ACNS}, 2018. [Online].
  Available: \url{https://doi.org/10.1007/978-3-319-93387-0_5}
\BIBentrySTDinterwordspacing

\bibitem[Genkin et~al.(2023)Genkin, Kosasih, Liu, Trikalinou, Unterluggauer,
  and Yarom]{GenkinKLTUY23}
D.~Genkin, W.~Kosasih, F.~Liu, A.~Trikalinou, T.~Unterluggauer, and Y.~Yarom,
  ``{CacheFX}: A framework for evaluating cache security,'' in \emph{AsiaCCS},
  2023.

\bibitem[Gerlach et~al.(2023)Gerlach, Schwarz, Faro{\ss}, and
  Schwarz]{gerlachEfficientGenericMicroarchitectural2023}
L.~Gerlach, S.~Schwarz, N.~Faro{\ss}, and M.~Schwarz, ``Efficient and generic
  microarchitectural hash-function recovery,'' in \emph{IEEE SP}, 2023.

\bibitem[Gruss et~al.(2015)Gruss, Spreitzer, and
  Mangard]{grussCacheTemplateAttacks2015}
\BIBentryALTinterwordspacing
D.~Gruss, R.~Spreitzer, and S.~Mangard, ``Cache template attacks: Automating
  attacks on inclusive last-level caches,'' in \emph{USENIX Sec.}, 2015.
  [Online]. Available:
  \url{https://www.usenix.org/conference/usenixsecurity15/technical-sessions/presentation/gruss}
\BIBentrySTDinterwordspacing

\bibitem[Gruss et~al.(2016{\natexlab{b}})Gruss, Maurice, and
  Mangard]{10.1007/978-3-319-40667-1_15}
D.~Gruss, C.~Maurice, and S.~Mangard, ``Rowhammer.js: A remote software-induced
  fault attack in {Javascript},'' in \emph{DIMVA}, 2016.

\bibitem[Gruss et~al.(2016{\natexlab{a}})Gruss, Maurice, Wagner, and
  Mangard]{grussFlushFlushFast2016}
D.~Gruss, C.~Maurice, K.~Wagner, and S.~Mangard, ``{Flush+Flush}: A fast and
  stealthy cache attack,'' in \emph{DIMVA}, 2016.

\bibitem[Horowitz et~al.(2024)Horowitz, Ronen, and
  Yarom]{horowitzSpecoScopeCacheProbing2024}
G.~Horowitz, E.~Ronen, and Y.~Yarom, ``{Spec-o-Scope}: Cache probing at cache
  speed,'' in \emph{CCS}, 2024.

\bibitem[Hund et~al.(2013)Hund, Willems, and Holz]{hundPracticalTimingSide2013}
R.~Hund, C.~Willems, and T.~Holz, ``Practical timing side channel attacks
  against kernel space {ASLR},'' in \emph{IEEE SP}, 2013.

\bibitem[Inci et~al.(2015)Inci, Gulmezoglu, Irazoqui, Eisenbarth, and
  Sunar]{inciSeriouslyGetMy2015}
\BIBentryALTinterwordspacing
M.~S. Inci, B.~Gulmezoglu, G.~Irazoqui, T.~Eisenbarth, and B.~Sunar,
  ``Seriously, get off my cloud! cross-{VM} {RSA} key recovery in a public
  cloud,'' 2015. [Online]. Available: \url{https://eprint.iacr.org/2015/898}
\BIBentrySTDinterwordspacing

\bibitem[{Intel}({\natexlab{a}})]{Intel64IA32}
{Intel}, ``Intel 64 and {IA}-32 architectures software developer manuals,''
  \url{https://www.intel.com/content/www/us/en/developer/articles/technical/intel-sdm.html}.

\bibitem[{Intel}({\natexlab{b}})]{intel12thGenerationIntel}
{Intel}, ``12th generation {Intel} {Core} processors,''
  \url{https://edc.intel.com/content/www/us/en/design/ipla/software-development-platforms/client/platforms/alder-lake-desktop/12th-generation-intel-core-processors-datasheet-volume-1-of-2/010/}.

\bibitem[{Intel}(2016)]{intel6thGenerationIntel2016}
{Intel}, ``6th generation {Intel} core processor family uncore performance
  monitoring reference manual,''
  \url{https://www.intel.com/content/www/us/en/content-details/671274/6th-generation-intel-core-processor-family-uncore-performance-monitoring-reference-manual.html},
  2016.

\bibitem[{Intel}({\natexlab{c}})]{intelIntelXeonProcessor}
{Intel}, ``Intel {Xeon} processor scalable memory family uncore performance
  monitoring reference manual,''
  \url{https://www.intel.com/content/www/us/en/content-details/671389/intel-xeon-processor-scalable-memory-family-uncore-performance-monitoring-reference-manual.html}.

\bibitem[Irazoqui et~al.(2015{\natexlab{a}})Irazoqui, Eisenbarth, and
  Sunar]{irazoquiSharedCacheAttack2015}
G.~Irazoqui, T.~Eisenbarth, and B.~Sunar, ``{S\$A}: A shared cache attack that
  works across cores and defies {VM} sandboxing -- and its application to
  {AES},'' in \emph{IEEE SP}, 2015.

\bibitem[Irazoqui et~al.(2015{\natexlab{b}})Irazoqui, Eisenbarth, and
  Sunar]{irazoquiSystematicReverseEngineering2015}
G.~Irazoqui, T.~Eisenbarth, and B.~Sunar, ``Systematic reverse engineering of
  cache slice selection in {Intel} processors,'' in \emph{DSD}, 2015.

\bibitem[Kang et~al.(2024)Kang, Wang, Kim, van Schaik, Tobah, Genkin, Kwong,
  and Yarom]{294601}
\BIBentryALTinterwordspacing
I.~Kang, W.~Wang, J.~Kim, S.~van Schaik, Y.~Tobah, D.~Genkin, A.~Kwong, and
  Y.~Yarom, ``{SledgeHammer}: Amplifying {Rowhammer} via bank-level
  parallelism,'' in \emph{USENIX Sec.}, 2024. [Online]. Available:
  \url{https://www.usenix.org/conference/usenixsecurity24/presentation/kang}
\BIBentrySTDinterwordspacing

\bibitem[Kaplan(2023)]{kaplanOptimizationAmplificationCache2023}
\BIBentryALTinterwordspacing
D.~A. Kaplan, ``Optimization and amplification of cache side channel signals,''
  arXiv 2303.00122, 2023. [Online]. Available:
  \url{http://arxiv.org/abs/2303.00122}
\BIBentrySTDinterwordspacing

\bibitem[Kario(2023)]{cryptoeprint:2023/1441}
\BIBentryALTinterwordspacing
H.~Kario, ``{Out of the Box Testing},'' Cryptology {ePrint} Archive, Paper
  2023/1441, 2023. [Online]. Available: \url{https://eprint.iacr.org/2023/1441}
\BIBentrySTDinterwordspacing

\bibitem[Katzman et~al.(2023)Katzman, Kosasih, Chuengsatiansup, Ronen, and
  Yarom]{katzmanGatesTimeImproving2023}
\BIBentryALTinterwordspacing
D.~Katzman, W.~Kosasih, C.~Chuengsatiansup, E.~Ronen, and Y.~Yarom, ``The gates
  of time: Improving cache attacks with transient execution,'' in \emph{USENIX
  Sec.}, 2023. [Online]. Available:
  \url{https://www.usenix.org/conference/usenixsecurity23/presentation/katzman}
\BIBentrySTDinterwordspacing

\bibitem[Kessous(2023)]{kessousPrunePlumTreeFindingEvictionSetsatScale2023}
T.~Kessous, ``{Prune-PlumTree}---finding-eviction-sets-at-scale,''
  \url{https://github.com/TomKessous/Prune-PlumTree---Finding-Eviction-Sets-at-Scale},
  2023.

\bibitem[Kessous and Gilboa(2024)]{kessousPrune+PlumTreeFindingEviction2024}
T.~Kessous and N.~Gilboa, ``{Prune+PlumTree} - finding eviction sets at
  scale,'' in \emph{IEEE SP}, 2024.

\bibitem[King(2024)]{kingColinIanKingStressng2024}
C.~I. King, ``{{Stress-Ng}},'' \url{https://github.com/ColinIanKing/stress-ng},
  2024.

\bibitem[Kocher et~al.(2019)Kocher, Horn, Fogh, Genkin, Gruss, Haas, Hamburg,
  Lipp, Mangard, Prescher, Schwarz, and
  Yarom]{kocherSpectreAttacksExploiting2019}
P.~Kocher, J.~Horn, A.~Fogh, D.~Genkin, D.~Gruss, W.~Haas, M.~Hamburg, M.~Lipp,
  S.~Mangard, T.~Prescher, M.~Schwarz, and Y.~Yarom, ``Spectre attacks:
  Exploiting speculative execution,'' in \emph{IEEE SP}, 2019.

\bibitem[Lipp et~al.(2018)Lipp, Schwarz, Gruss, Prescher, Haas, Fogh, Horn,
  Mangard, Kocher, Genkin, Yarom, and Hamburg]{lippMeltdownReadingKernel2018}
\BIBentryALTinterwordspacing
M.~Lipp, M.~Schwarz, D.~Gruss, T.~Prescher, W.~Haas, A.~Fogh, J.~Horn,
  S.~Mangard, P.~Kocher, D.~Genkin, Y.~Yarom, and M.~Hamburg, ``Meltdown:
  Reading kernel memory from user space,'' in \emph{USENIX Sec.}, 2018.
  [Online]. Available:
  \url{https://www.usenix.org/conference/usenixsecurity18/presentation/lipp}
\BIBentrySTDinterwordspacing

\bibitem[Liu et~al.(2015)Liu, Yarom, Ge, Heiser, and
  Lee]{liuLastLevelCacheSideChannel2015}
F.~Liu, Y.~Yarom, Q.~Ge, G.~Heiser, and R.~B. Lee, ``Last-level cache
  side-channel attacks are practical,'' in \emph{IEEE SP}, 2015.

\bibitem[Maurice et~al.(2015)Maurice, Le~Scouarnec, Neumann, Heen, and
  Francillon]{mauriceReverseEngineeringIntel2015}
C.~Maurice, N.~Le~Scouarnec, C.~Neumann, O.~Heen, and A.~Francillon, ``Reverse
  engineering {Intel} last-level cache complex addressing using performance
  counters,'' in \emph{RAID}, 2015.

\bibitem[McCalpin(2021)]{mccalpinMappingAddressesL32021}
J.~D. McCalpin, ``Mapping addresses to {L3/CHA} slices in {Intel} processors,''
  \url{https://hdl.handle.net/2152/87595}, 2021.

\bibitem[Morgan(2024)]{morganBMorgan1296Perfcounters2024}
B.~Morgan, ``{BMorgan1296/Perfcounters},''
  \url{https://github.com/BMorgan1296/perfcounters}, 2024.

\bibitem[O'Connell et~al.(2024)O'Connell, Sour, Magen, Genkin, Oren, Shacham,
  and Yarom]{294651}
\BIBentryALTinterwordspacing
S.~O'Connell, L.~A. Sour, R.~Magen, D.~Genkin, Y.~Oren, H.~Shacham, and
  Y.~Yarom, ``Pixel thief: Exploiting {SVG} filter leakage in {Firefox} and
  {Chrome},'' in \emph{USENIX Sec.}, 2024. [Online]. Available:
  \url{https://www.usenix.org/conference/usenixsecurity24/presentation/oconnell}
\BIBentrySTDinterwordspacing

\bibitem[Oren et~al.(2015)Oren, Kemerlis, Sethumadhavan, and
  Keromytis]{orenSpySandboxPractical2015}
Y.~Oren, V.~P. Kemerlis, S.~Sethumadhavan, and A.~D. Keromytis, ``The spy in
  the sandbox: Practical cache attacks in {JavaScript} and their
  implications,'' in \emph{CCS}, 2015.

\bibitem[Osvik et~al.(2005)Osvik, Shamir, and
  Tromer]{osvikCacheAttacksCountermeasures2005}
\BIBentryALTinterwordspacing
D.~A. Osvik, A.~Shamir, and E.~Tromer, ``Cache attacks and countermeasures: The
  case of {AES},'' IACR ePrint archive 2005/271, 2005. [Online]. Available:
  \url{https://eprint.iacr.org/2005/271}
\BIBentrySTDinterwordspacing

\bibitem[Paccagnella et~al.(2021)Paccagnella, Luo, and
  Fletcher]{paccagnellaLordRingSide}
\BIBentryALTinterwordspacing
R.~Paccagnella, L.~Luo, and C.~W. Fletcher, ``Lord of the ring(s): Side channel
  attacks on the {CPU} on-chip ring interconnect are practical,'' in
  \emph{USENIX Sec.}, 2021. [Online]. Available:
  \url{https://www.usenix.org/conference/usenixsecurity21/presentation/paccagnella}
\BIBentrySTDinterwordspacing

\bibitem[Percival(2005)]{percivalCacheMissingFun2005}
\BIBentryALTinterwordspacing
C.~Percival, ``Cache missing for fun and profit,'' in \emph{BSDCan}, 2005.
  [Online]. Available: \url{https://www.daemonology.net/papers/htt.pdf}
\BIBentrySTDinterwordspacing

\bibitem[Purnal et~al.(2021{\natexlab{b}})Purnal, Giner, Gruss, and
  Verbauwhede]{PurnalGGV21}
A.~Purnal, L.~Giner, D.~Gruss, and I.~Verbauwhede, ``Systematic analysis of
  randomization-based protected cache architectures,'' in \emph{{SP}}, 2021.

\bibitem[Purnal et~al.(2021{\natexlab{a}})Purnal, Turan, and
  Verbauwhede]{purnalPrimeScopeOvercoming2021}
A.~Purnal, F.~Turan, and I.~Verbauwhede, ``{Prime+Scope}: Overcoming the
  observer effect for high-precision cache contention attacks,'' in \emph{CCS},
  2021.

\bibitem[Purnal et~al.(2023)Purnal, Bognar, Piessens, and
  Verbauwhede]{PurnalBPV23}
A.~Purnal, M.~Bognar, F.~Piessens, and I.~Verbauwhede, ``{ShowTime}: Amplifying
  arbitrary {CPU} timing side channels,'' in \emph{AsiaCCS}, 2023.

\bibitem[Quinlan(1986)]{10.1007/BF00116251}
J.~R. Quinlan, ``Induction of decision trees,'' in \emph{Machine Learning},
  1986.

\bibitem[Qureshi(2018)]{Qureshi18}
M.~K. Qureshi, ``{CEASER:} mitigating conflict-based cache attacks via
  encrypted-address and remapping,'' in \emph{{MICRO}}, 2018.

\bibitem[Shusterman et~al.(2021)Shusterman, Agarwal, O'Connell, Genkin, Oren,
  and Yarom]{272258}
\BIBentryALTinterwordspacing
A.~Shusterman, A.~Agarwal, S.~O'Connell, D.~Genkin, Y.~Oren, and Y.~Yarom,
  ``Prime+{Probe}~1, {JavaScript}~0: Overcoming browser-based side-channel
  defenses,'' in \emph{USENIX Sec.}, 2021. [Online]. Available:
  \url{https://www.usenix.org/conference/usenixsecurity21/presentation/shusterman}
\BIBentrySTDinterwordspacing

\bibitem[Song and Liu(2019)]{songDynamicallyFindingMinimal2019}
\BIBentryALTinterwordspacing
W.~Song and P.~Liu, ``Dynamically finding minimal eviction sets can be quicker
  than you think for side-channel attacks against the {LLC},'' in \emph{RAID},
  2019. [Online]. Available:
  \url{https://www.usenix.org/conference/raid2019/presentation/song}
\BIBentrySTDinterwordspacing

\bibitem[Thoma and G{\"{u}}neysu(2022)]{ThomaG22}
J.~P. Thoma and T.~G{\"{u}}neysu, ``Write me and {I'll} tell you secrets -
  write-after-write effects on {Intel} {CPUs},'' in \emph{{RAID}}, 2022.

\bibitem[Tsunoo et~al.(2002)Tsunoo, Tsujihara, Minematsu, and
  Miyauchi]{tsunooCryptanalysisBlockCiphers2002}
Y.~Tsunoo, E.~Tsujihara, K.~Minematsu, and H.~Miyauchi, ``Cryptanalysis of
  block ciphers implemented on computers with cache,'' in \emph{ISITA}, 2002.

\bibitem[Tsunoo et~al.(2003)Tsunoo, Saito, Suzaki, Shigeri, and
  Miyauchi]{tsunooCryptanalysisImplementedComputers2003}
Y.~Tsunoo, T.~Saito, T.~Suzaki, M.~Shigeri, and H.~Miyauchi, ``Cryptanalysis of
  {DES} implemented on computers with cache,'' in \emph{CHES}, 2003.

\bibitem[Vanhoef and Ronen(2020)]{9152782}
M.~Vanhoef and E.~Ronen, ``Dragonblood: Analyzing the {Dragonfly} handshake of
  {WPA3} and {EAP-pwd},'' in \emph{IEEE SP}, 2020.

\bibitem[Vila et~al.(2019)Vila, K{\"o}pf, and
  Morales]{vilaTheoryPracticeFinding2019}
P.~Vila, B.~K{\"o}pf, and J.~F. Morales, ``Theory and practice of finding
  eviction sets,'' in \emph{IEEE SP}, 2019.

\bibitem[{Vimalm} and McCalpin(2022)]{vimalmReTurningCache2022}
\BIBentryALTinterwordspacing
{Vimalm} and J.~D. McCalpin, ``Re: Turning off the cache coherence directory
  system wide in {Intel Xeon Gold} 6242 processor,'' 2022. [Online]. Available:
  \url{https://community.intel.com/t5/Processors/Turning-off-the-cache-coherence-directory-system-wide-in-Intel/m-p/1364445#M56606}
\BIBentrySTDinterwordspacing

\bibitem[Wang et~al.(2024)Wang, Paccagnella, Wahby, and Brown]{WangPWB24}
P.-L. Wang, R.~Paccagnella, R.~S. Wahby, and F.~Brown, ``Bending
  microarchitectural weird machines towards practicality,'' in \emph{{USENIX}
  Sec.}, 2024.

\bibitem[Werner et~al.(2019)Werner, Unterluggauer, Giner, Schwarz, Gruss, and
  Mangard]{WernerUG0GM19}
M.~Werner, T.~Unterluggauer, L.~Giner, M.~Schwarz, D.~Gruss, and S.~Mangard,
  ``{ScatterCache}: Thwarting cache attacks via cache set randomization,'' in
  \emph{{USENIX} Sec.}, 2019.

\bibitem[Wu et~al.(2012)Wu, Xu, and Wang]{WuXW12}
Z.~Wu, Z.~Xu, and H.~Wang, ``Whispers in the hyper-space: High-speed covert
  channel attacks in the cloud,'' in \emph{{USENIX} Sec.}, 2012.

\bibitem[Xue et~al.(2023)Xue, Han, and Song]{xueCTPPFastStealth2023}
Z.~Xue, J.~Han, and W.~Song, ``{CTPP}: A fast and stealth algorithm for
  searching eviction sets on {Intel} processors,'' in \emph{RAID}, 2023.

\bibitem[Yan et~al.(2019)Yan, Sprabery, Gopireddy, Fletcher, Campbell, and
  Torrellas]{yanAttackDirectoriesNot2019}
M.~Yan, R.~Sprabery, B.~Gopireddy, C.~Fletcher, R.~Campbell, and J.~Torrellas,
  ``Attack directories, not caches: Side channel attacks in a non-inclusive
  world,'' in \emph{IEEE SP}, 2019.

\bibitem[Yan et~al.(2020)Yan, Fletcher, and Torrellas]{244042}
\BIBentryALTinterwordspacing
M.~Yan, C.~W. Fletcher, and J.~Torrellas, ``Cache telepathy: Leveraging shared
  resource attacks to learn {DNN} architectures,'' in \emph{USENIX Sec.}, 2020.
  [Online]. Available:
  \url{https://www.usenix.org/conference/usenixsecurity20/presentation/yan}
\BIBentrySTDinterwordspacing

\bibitem[Yarom and Falkner(2014)]{yaromFLUSHRELOADHigh2014}
\BIBentryALTinterwordspacing
Y.~Yarom and K.~Falkner, ``{Flush+Reload}: A high resolution, low noise, {L3}
  cache side-channel attack,'' in \emph{USENIX Sec.}, 2014. [Online].
  Available:
  \url{https://www.usenix.org/conference/usenixsecurity14/technical-sessions/presentation/yarom}
\BIBentrySTDinterwordspacing

\bibitem[Yarom et~al.(2015)Yarom, Ge, Liu, Lee, and
  Heiser]{yaromMappingIntelLastLevel2015}
\BIBentryALTinterwordspacing
Y.~Yarom, Q.~Ge, F.~Liu, R.~B. Lee, and G.~Heiser, ``Mapping the {Intel}
  last-level cache,'' IACR ePrint 2015/905, 2015. [Online]. Available:
  \url{https://eprint.iacr.org/2015/905}
\BIBentrySTDinterwordspacing

\bibitem[Younis et~al.(2015)Younis, Kifayat, Shi, and
  Askwith]{younisNewPrimeProbe2015}
Y.~A. Younis, K.~Kifayat, Q.~Shi, and B.~Askwith, ``A new prime and probe cache
  side-channel attack for cloud computing,'' in \emph{CIT}, 2015.

\bibitem[Yuan et~al.(2022)Yuan, Pang, and Wang]{277232}
\BIBentryALTinterwordspacing
Y.~Yuan, Q.~Pang, and S.~Wang, ``Automated side channel analysis of media
  software with manifold learning,'' in \emph{USENIX Sec.}, 2022. [Online].
  Available:
  \url{https://www.usenix.org/conference/usenixsecurity22/presentation/yuan-yuanyuan}
\BIBentrySTDinterwordspacing

\bibitem[Zhao et~al.(2024)Zhao, Morrison, Fletcher, and
  Torrellas]{zhaoLastLevelCacheSideChannel2024}
Z.~N. Zhao, A.~Morrison, C.~W. Fletcher, and J.~Torrellas, ``Last-level cache
  side-channel attacks are feasible in the modern public cloud,'' in
  \emph{ASPLOS}, 2024.

\end{thebibliography}
\algnewcommand{\algorithmicgoto}{\textbf{Go to}}
\algnewcommand{\Goto}[1]{\algorithmicgoto~\ref{#1}}

\renewcommand{\thesection}{\Alph{section}}
\appendices
\crefalias{section}{appendix}

\section{Comparator Gate Code}\label{app:comparator-gate}
\cref{lst:comparator-gate} shows the assembly code for the comparator gate, which is used to compare memory latencies for two cache lines (\texttt{input} and \texttt{compare}) in the LLC\@.
The code is just-in-time compiled using AssemblyLine~\cite{chuengsatiansup0xADE1A1DEAssemblyLine2022} for ease of use.
This is the main building block of our userspace slice index prediction algorithm.

\parhead{Running a Single Measurement.}
To carry out a single measurement using the comparator gate:
\begin{itemize}
    \item \textbf{Setup:} Place the \texttt{input}, \texttt{compare} and \texttt{signal} cache lines into the LLC using L2 eviction sets.
    \item \textbf{Lines 1--11:} Halt speculation with memory fences, place \texttt{signal} in R10 for later use as RDTSCP will overwrite RDX\@.
    \item \textbf{Lines 19--22:} Set up a speculative window with return-based gadget~\cite{kaplanOptimizationAmplificationCache2023}, squashing the transient branch when the request for \texttt{input} in RDI completes.
    \item \textbf{Lines 13--15:} Access \texttt{compare} in RSI, followed by a hand-tuned fixed-length delay chain of instructions as in~\cite{katzmanGatesTimeImproving2023}, to prevent the processor from accessing \texttt{signal} before the \texttt{compare} cache line arrives.
    \item \textbf{Lines 16--17:} Access \texttt{signal} in R10 after the chain and halt speculation with a load fence.
    \item \textbf{Lines 24--35:} Determine whether \texttt{signal} is cached in the L1 or not using the RDTSCP timer instruction.
\end{itemize}

\begin{lstlisting}[language=nasm,style=nasm, caption={Comparator Gate Assembly Code}, float, label={lst:comparator-gate}]
xor rdi, 0x800
mov rax, [rdi] ; Cache *input's page in TLB
xor rdi, 0x800
xor rsi, 0x800
mov rax, [rsi] ; Cache *compare's page in TLB
xor rsi, 0x800
mov r10, rdx
mfence
lfence
; DC_LEN sets the delay chain length
call long 0x9 + (DC_LEN * 0x3) ; Call L19

mov rsi, [rsi] ; Access *compare
add r10, rsi
...            ; Hand-tuned delay chain
mov r10, [r10] ; Access *signal
lfence         ; Halt speculation

mov rax, 0x16 + (DC_LEN * 0x3) ; Offset to L24
add rax, [rdi] ; Access *input
add [rsp], rax ; Set return address to L24
ret            ; Mispredict to L13

lfence         ; Halt speculation
xor r10, 0x800
mov rax, [r10] ; Cache *signal's page in TLB
xor r10, 0x800
mfence
lfence
rdtscp
mov r9, rax
add r10, [r10] ; Measure access to *signal
rdtscp
sub rax, r9
ret            ; Return access time
\end{lstlisting}

\section{Generic Slice Function Retrieval}\label{app:slice-function-retrieval-algorithms}
To build the decision tree and improve our slice predictions, we need to first recover the slice function in a generic and time-efficient manner for any modern Intel processor.
In our threat model, the attacker obtains the targeted processor and recovers the function offline.
This only needs to be completed once per processor.

\parhead{Slice Function Recovery Methods.}
\cref{tab:slice_recovery_comparison} details prior methods for reverse engineering the slice function, comparing the method of slice recovery for an address, the entire function recovery, and time taken.
The majority of prior techniques are manual, and none beside work by \citet{gerlachEfficientGenericMicroarchitectural2023} can automatically recover the function in linear and non-linear cases.
We aim to improve on this by providing an efficient and automated method for both linear and non-linear functions.
Our codebase details the specifics of the implementation, but we provide a high-level overview here.

\begin{table}[!t]
\caption{Recovery methods for the Intel LLC function.}
\begin{adjustbox}{max width=\linewidth}
\begin{tabular}{llllll}
\toprule                                                 &                                                                           &                                                                                                & \multicolumn{2}{c}{\textbf{Retrieval Time}}                                        \\ \cmidrule(lr){4-5}
\textbf{Work}                                            & \begin{tabular}[c]{@{}l@{}}\textbf{Slice}\\\textbf{Recovery}\end{tabular} & \begin{tabular}[c]{@{}l@{}}\textbf{Function}\\\textbf{Recovery}\end{tabular}                   & \textbf{Linear}                                            & \textbf{\begin{tabular}[c]{@{}l@{}}Non-\\ Linear\end{tabular}}   \\ \hline \citet{hundPracticalTimingSide2013}                      & Cache Conflict                                                            & Manual                                                                                         & Manual                                                     & --                    \\ \citet{irazoquiSystematicReverseEngineering2015}         & Cache Conflict                                                            & Linear Eq.                                                                                     & Manual                                                     & --                    \\ \citet{yaromMappingIntelLastLevel2015}                   & Access Timing                                                             & Manual                                                                                         & --                                                         & Manual                \\ \citet{inciSeriouslyGetMy2015}                           & Cache Conflict                                                            & \begin{tabular}[c]{@{}l@{}}Manual\end{tabular}                                                 & --                                                         & Manual                \\
\citet{mccalpinMappingAddressesL32021}                   & Perf. Counter                                                             & Manual                                                                                         & Manual                                                     & Manual                \\
\citet{mauriceReverseEngineeringIntel2015}               & Perf. Counter                                                             & \begin{tabular}[c]{@{}l@{}}Automated \\ Grey-box\end{tabular}                                  & Minutes                                                    & --                    \\
\citet{gerlachEfficientGenericMicroarchitectural2023}    & \begin{tabular}[c]{@{}l@{}}Perf. Counter, \\ Access Timing\end{tabular}   & \begin{tabular}[c]{@{}l@{}}Automated \\ Black-box\end{tabular}                                 & Minutes                                                    & Weeks                 \\
Ours                                                     & Perf. Counter                                                             & \begin{tabular}[c]{@{}l@{}}Direct physical \\ address access\end{tabular}                 & $\sim$25 ms                                      & \textless{}1000 ms   \\
\bottomrule
\end{tabular}
\end{adjustbox}
\label{tab:slice_recovery_comparison}
\end{table}

\parhead{Determining Memory Slice Index Using Performance Counters.}
To retrieve any information regarding the slice function, the first step is to identify which slice a memory address maps to.

The known reverse engineering process requires finding two memory addresses differing by a single physical bit to recover slice information~\cite{mauriceReverseEngineeringIntel2015, mccalpinMappingAddressesL32021}.
This approach requires searching for pairs of addresses in large buffers of memory and is limited to the currently installed system RAM\@.

We implement a custom kernel module which bypasses memory controller checks, enabling read requests for any physical address of our choosing without the need for a large buffer of memory.
By directly accessing any physical memory address, we can reconstruct the slice function up to the memory limit of the processor.

We likewise use the CLFLUSH instruction to generate read requests and observe slice lookups~\cite{mauriceReverseEngineeringIntel2015, mccalpinMappingAddressesL32021}.
To interface with the uncore performance counters~\cite{Intel64IA32, intel6thGenerationIntel2016}, we employ a custom library~\cite{morganBMorgan1296Perfcounters2024} and monitor read lookups\footnote{Using \texttt{UNC\_CBO\_CACHE\_LOOKUP.ANY\_MESI} perf counter.} on each available slice.
By repeatedly issuing CLFLUSH commands on an address, we can identify the slice to which an address maps by observing the performance counter with the highest number of read requests.

\begin{table}[t!]
\caption{Recovery times for slice functions. Non-linear take longer due to the sequence-centric recovery approach.}
\centering
\begin{tabularx}{\linewidth}{l>{\raggedleft\arraybackslash}X>{\raggedleft\arraybackslash}X>{\raggedleft\arraybackslash}X}
\toprule
\multicolumn{1}{l}{\textbf{Processor}} & \multicolumn{1}{r}{\textbf{Slices}} & \multicolumn{1}{r}{\textbf{Linear}} & \multicolumn{1}{r}{\textbf{Time (ms)}} \\
\midrule
i7-6700K & 4 & \cmark & 25 \\
i7-11700KF & 8 & \cmark & 19 \\
i7-13700H & 8 & \cmark & 24 \\
i7-8700 & 6 & \xmark & 291 \\
i7-9850H & 6 & \xmark & 225 \\
i7-10710U & 6 & \xmark & 190 \\
i9-12900KF & 10 & \xmark & 522 \\
i9-13900KF & 12 & \xmark & 740 \\
i9-14900K & 12 & \xmark & 930 \\
\bottomrule
\end{tabularx}
\label{tab:slice_recovery_time}
\end{table}

\begin{table}[h]
\centering
\caption{Recovered slice functions for various Intel Core processor generations with a power of two slice count. The XOR permutation masks \textit{m} are used to determine each slice index bit.}
\begin{adjustbox}{width=\linewidth}
\begin{tabular}{lllr}
\toprule
\textbf{Year} & \textbf{Generation} & \textbf{Slices} & \textbf{Permutation Masks} \\
\midrule
2017                              & 7                        & 2                  & $m_0 = \texttt{0x5b5f575440}$ \\ \\
\multirow{2}{*}{\begin{tabular}[c]{@{}l@{}}2013, 2015,\\ 2017\end{tabular}} & \multirow{2}{*}{4--7} & \multirow{2}{*}{4} & $m_0 = \texttt{0x5b5f575440}$ \\
                                  &                          &                    & $m_1 = \texttt{0x6eb5faa880}$ \\ \\
\multirow{3}{*}{2021, 2023}       & \multirow{3}{*}{11, 13}  & \multirow{3}{*}{8} & $m_0 = \texttt{0x5b5f575440}$ \\
                                  &                          &                    & $m_1 = \texttt{0x71aeeb1200}$ \\
                                  &                          &                    & $m_2 = \texttt{0x06d87f2c00}$ \\
\bottomrule
\end{tabular}
\end{adjustbox}
\label{tab:intel_processors_pot}
\end{table}

\begin{table}[h]
\centering
\caption{Recovered slice functions for various Intel Core processor generations with a non-power of two slice count. The XOR permutation masks \textit{m} calculate each index bit for a lookup into the base sequences shown in \cref{tab:intel_processors_npot_base_sequence}.}
\begin{adjustbox}{width=\linewidth}
\begin{tabular}{lllr}
\toprule
\textbf{Year} & \textbf{Generation} & \textbf{Slices} & \textbf{Permutation Masks} \\
\midrule
\multirow{7}{*}{\begin{tabular}[c]{@{}c@{}}2018, 2019,\\2020, 2021\end{tabular}} & \multirow{7}{*}{\begin{tabular}[c]{@{}c@{}}8, 9, 10, 12\end{tabular}} & \multirow{7}{*}{6, 10} & $m_0 = \texttt{0x21ae7be000}$ \\
                                              &                                   &                            & $m_1 = \texttt{0x435cf7c000}$ \\
                                              &                                   &                            & $m_2 = \texttt{0x2717946000}$ \\
                                              &                                   &                            & $m_3 = \texttt{0x4e2f28c000}$ \\
                                              &                                   &                            & $m_4 = \texttt{0x1c5e518000}$ \\
                                              &                                   &                            & $m_5 = \texttt{0x38bca30000}$ \\
                                              &                                   &                            & $m_6 = \texttt{0x50d73de000}$ \\ \\
\multirow{7}{*}{\begin{tabular}[c]{@{}c@{}}2022\end{tabular}} & \multirow{7}{*}{\begin{tabular}[c]{@{}c@{}}13\end{tabular}} & \multirow{7}{*}{12} & $m_0 = \texttt{0x52c6a78000}$ \\
                                              &                                   &                            & $m_1 = \texttt{0x30342b8000}$ \\
                                              &                                   &                            & $m_2 = \texttt{0x547f480000}$ \\
                                              &                                   &                            & $m_3 = \texttt{0x3d47f48000}$ \\
                                              &                                   &                            & $m_4 = \texttt{0x1c5e518000}$ \\
                                              &                                   &                            & $m_5 = \texttt{0x38bca30000}$ \\
                                              &                                   &                            & $m_6 = \texttt{0x23bfe18000}$ \\
                                              &                                   &                            & $m_7 = \texttt{0x0000000000}$ \\
                                              &                                   &                            & $m_8 = \texttt{0x7368dc0000}$ \\ \\
\multirow{7}{*}{\begin{tabular}[c]{@{}c@{}}2023\end{tabular}} & \multirow{7}{*}{\begin{tabular}[c]{@{}c@{}}14\end{tabular}} & \multirow{7}{*}{12} & $m_0 = \texttt{0x2f52c6a78000}$ \\
                                              &                                   &                            & $m_1 = \texttt{0x0cb0342b8000}$ \\
                                              &                                   &                            & $m_2 = \texttt{0x35d47f480000}$ \\
                                              &                                   &                            & $m_3 = \texttt{0x39bd47f48000}$ \\
                                              &                                   &                            & $m_4 = \texttt{0x109c5e518000}$ \\
                                              &                                   &                            & $m_5 = \texttt{0x2038bca30000}$ \\
                                              &                                   &                            & $m_6 = \texttt{0x0e23bfe18000}$ \\
                                              &                                   &                            & $m_7 = \texttt{0x000000000000}$ \\
                                              &                                   &                            & $m_8 = \texttt{0x31f368dc0000}$ \\
\bottomrule
\end{tabular}
\end{adjustbox}
\label{tab:intel_processors_npot}
\end{table}

\begin{table}[h]
\caption{Intel Core processor generations with non-power-of-two slices and their corresponding base sequences.}
\begin{adjustbox}{max width=\linewidth}
\begin{tabular}{rrrl}
\toprule
\textbf{Year} & \textbf{Generation} & \textbf{Slices} & \multicolumn{1}{c}{\textbf{Base Sequence}} \\
\midrule
2018, 2019    & 8, 9                & 6               & \begin{tabular}[c]{@{}l@{}}\texttt{0123 1434 1032 0525 1032 0525 0525 1434} \\ \texttt{0123 5052 5052 4143 1032 4143 4143 5052} \\ \texttt{2301 5250 3210 4341 3210 4341 4341 5250} \\ \texttt{2301 3414 3414 2505 3210 2505 2505 3414}\end{tabular} \\ \\
2020, 2021    & 10, 12              & 10              & \begin{tabular}[c]{@{}l@{}}\texttt{0505 3636 1414 2727 1414 2727 0589 3698} \\ \texttt{4141 7272 5098 6389 5050 6363 4189 7298} \\ \texttt{4141 7272 5050 6363 5050 6363 8941 9872} \\ \texttt{0505 3636 9814 8927 1414 2727 8905 9836}\end{tabular} \\ \\
2022, 2023    & 13, 14              & 12              & \begin{tabular}[c]{@{}l@{}}\texttt{0145 1858 3276 2b6b 1054 0949 2b6b 3a7a} \\ \texttt{2367 b2b6 9094 8185 3276 a3a7 8185 9094} \\ \texttt{6723 b6b2 5410 8581 7632 a7a3 8581 9490} \\ \texttt{4501 5818 7a3a 6b2b 5410 4909 6b2b 7a3a} \\ \texttt{6b2b 7632 5818 4501 7a3a 6723 4909 5818} \\ \texttt{8581 5410 b6b2 a7a3 9490 4501 a7a3 b6b2} \\ \texttt{8185 1054 b2b6 2367 9094 0145 a3a7 b2b6} \\ \texttt{2b6b 3276 1858 0949 3a7a 2367 0949 1858} \\ \texttt{3a7a 2367 0949 1054 2b6b 3a7a 1858 0145} \\ \texttt{9094 8185 a3a7 3276 8185 9094 b2b6 2367} \\ \texttt{9490 4501 a7a3 7632 8581 9490 b6b2 6723} \\ \texttt{7a3a 6b2b 4909 5410 6b2b 7a3a 5818 4501} \\ \texttt{5410 4909 6723 7a3a 4909 5818 7632 6b2b} \\ \texttt{b6b2 a7a3 4501 9490 a7a3 b6b2 5410 8581} \\ \texttt{3276 a3a7 0145 9094 a3a7 b2b6 1054 8185} \\ \texttt{1858 0949 2367 3a7a 0949 1858 3276 2b6b}\end{tabular} \\
\bottomrule
\end{tabular}

\end{adjustbox}
\label{tab:intel_processors_npot_base_sequence}
\end{table}

\subsection{Linear Functions.}\label{subsec:linear-functions.}
Linear slice functions use a basic XOR operation with permutation masks to select address bits for calculating the slice index.
To recover the slice function, we analyse the slices for physical addresses \texttt{0x0}, \texttt{0x40}, \texttt{0x80}, and so forth, up to the maximum physical address supported by the processor.
We start with \texttt{0x40} instead of \texttt{0x0} because the lower bits (cache line offset bits) are not involved in the slice function.
For instance, let $H$ denote the unknown slice function and the recovered slice index as $s$, we have $s = H(\texttt{0x0}) \bigoplus H(\texttt{0x40})$.
We then decompose $s$ bit-wise and map it across the permutation masks array $m$ at index $\log_2(\texttt{0x40})=6$.
By recording the slice index for each tested physical address and mapping the bits of $s$ to the permutation masks array $m$, we can reconstruct the permutation masks bit by bit.
Repeating this process for each significant bit allows us to fully recover the slice function used by the processor.

\subsection{Non-Linear Functions.}\label{subsec:non-linear-functions.}
In contrast, recovering non-linear functions is inherently more complex.
To address this, we develop a method to simultaneously recover both the permutation masks and the base sequence.
Initially, we do not know the length of the sequence, so we start with an initial guess of length 1.
We measure slice mappings from address \texttt{0x0} to establish the initial base sequence $B$, again using performance counters and the kernel module.
We then generate a temporary sequence $T$ of slice mappings for the next address bit, comparing these mappings to our guessed base sequence by searching for the XOR permutation used.
Specifically, we search for the XOR permutation $x$ where $T \bigoplus x = B$.
If we find a matching $x$, then we split this bit-wise into the permutation masks array $m$ and continue to the next significant address bit, the same method as the linear function.
If we do not find any matching XOR permutations, we double the sequence length and repeat the process.
For the new length, we measure slice mappings from $0 \ldots 2^{n-1}$, where $n$ is the current guessed sequence length.
By iterating through this process—starting with a small sequence length and progressively increasing it until the correct length is found, we can reconstruct the permutation masks and the base sequence.

\cref{tab:slice_recovery_time} displays the time taken to extract the slice function on several generations of processors.
As can be seen, our automated recovery approach can retrieve this in under 30ms for linear and under a second for non-linear.

\subsection{Recovered Slice Functions}\label{app:slice_functions}
\cref{tab:intel_processors_pot,tab:intel_processors_npot} detail the recovered slice functions for various Intel Core processor generations.
\cref{tab:intel_processors_npot_base_sequence} shows the base sequences for non-power-of-two slice counts.

\newpage
\section{Meta-Review}
The following meta-review was prepared by the program committee for the 2025 IEEE Symposium on Security and Privacy (S\&P) as part of the review process as detailed in the call for papers.

\subsection{Summary}

This paper provides a number of techniques that jointly significantly improve the performance of generating last-level cache eviction sets. The paper first proposes a way to identify the slice that an address maps to by using a ``weird gate'' that performs a timing comparison. The paper also proposes methods for intra-page propagation of slice mappings. Based on these new proposals, the paper in the end succeeds in speeding up the eviction set generation for the full last-level-cache.

\subsection{Scientific Contributions}
\begin{itemize}

\item Creates a New Tool to Enable Future Science
\item Addresses a Long-Known Issue
\item Provides a Valuable Step Forward in an Established Field
\item Establishes a New Research Direction
\end{itemize}

\subsection{Reasons for Acceptance}
\begin{enumerate}

\item This paper creates a new tool to generate certain eviction sets. This helps to assess the security of CPU microarchitecture designs.

\item  This paper addresses a long-known issue, in that identification of eviction sets is a common prerequisite for cache-based attacks.

\item This paper provides a valuable step forward by significantly increasing the efficiency with which eviction sets can be identified.

\item This paper establishes a new research direction as the PC found some of the slice+slice technique to be novel.
\end{enumerate}

\subsection{Noteworthy Concerns}
\begin{enumerate}

\item The PC found this paper to be focused very deeply on a single aspect of a single vendor's CPU design. While the paper does a nice job building an effective eviction set identification tool out of access latency variations, it is ultimately a reactive piece of work focused on specific CPU makes and models.
\end{enumerate}

\end{document}